\documentclass[sigconf,nonacm]{acmart}

\definecolor{darkgreen}{HTML}{3C8031} 

\usepackage[commandnameprefix=always,final]{changes}
\definechangesauthor[name={R2}, color=blue]{R2}
\definechangesauthor[name={R3}, color=blue]{R3}
\definechangesauthor[name={1AC}, color=blue]{1AC}
\definechangesauthor[name={2AC}, color=blue]{2AC}

\usepackage{subcaption}
\usepackage{tabularx}
\usepackage{pifont}
\usepackage{float}
\usepackage{diagbox}
\usepackage{xcolor}
\usepackage[export]{adjustbox}
\usepackage{subcaption}
\usepackage{array}
\usepackage{pifont}

\newcolumntype{P}[1]{>{\centering\arraybackslash}p{#1}}

\AtBeginDocument{%
  \providecommand\BibTeX{{%
    \normalfont B\kern-0.5em{\scshape i\kern-0.25em b}\kern-0.8em\TeX}}}

\usepackage{todonotes}
\usepackage{makecell}
\usepackage{xcolor}

\newcommand{\system}{HappyRouting}

  \acmConference[MobileHCI '25]{MobileHCI '25: International Conference on Mobile Human-Computer Interaction}{September 22--September 26, 2025}{Sharm El-Sheikh, Egypt}
\acmBooktitle{MobileHCI '25: International Conference on Mobile Human-Computer Interaction,
  September 22--September 26, Sharm El-Sheikh, Egypt}
\begin{document}

%\title{Contextualizing in-the-wild emotion recognition with remote smartphone sensing feeds}
\title[HappyRouting]{HappyRouting: Learning Emotion-Aware Route Trajectories for Scalable In-The-Wild Navigation}
%\title{}
%Technical Design Space Analysis for Ubiquitous Driver Emotion Assessment Using Audio-Visual, Contextual, and Environmental Input

\author{David Bethge}
\affiliation{%
  \institution{Porsche AG, LMU Munich}
  \city{Stuttgart}
  \country{Germany}
}
\orcid{0000-0002-0031-0565}
\email{david.bethge@ifi.lmu.de}

\author{Daniel Bulanda}
\affiliation{%
  \institution{GrapeUp}
  \city{Warsaw}
  \country{Poland}
}
\author{Adam Kozlowski}
\affiliation{%
  \institution{GrapeUp}
  \city{Warsaw}
  \country{Poland}
}

\author{Thomas Kosch}
\affiliation{%
  \institution{HU Berlin}
  \city{Berlin}
  \country{Germany}
}
\orcid{0000-0001-6300-9035}

\author{Albrecht Schmidt}
\affiliation{%
  \institution{LMU Munich}
  \city{Munich}
  \country{Germany}
}
\orcid{0000-0003-3890-1990}

\author{Tobias Grosse-Puppendahl}
\affiliation{%
  \institution{Porsche AG}
  \city{Stuttgart}
  \country{Germany}
}
\orcid{0000-0003-4961-6554}

\renewcommand{\shortauthors}{Bethge et al.}


%Problem
%Solution
%Approach
%Findings
%Conclusion

\begin{abstract}

Routes represent an integral part of triggering emotions in drivers. Navigation systems allow users to choose a navigation strategy, such as the fastest or shortest route. However, they do not consider the driver's emotional well-being. We present HappyRouting, a novel navigation-based empathic car interface guiding drivers through real-world traffic while evoking positive emotions. We propose design considerations, derive a technical architecture, and implement a routing optimization framework. Our contribution is a machine learning-based generated emotion map layer, predicting emotions along routes based on static and dynamic contextual data. We evaluated HappyRouting in a real-world driving study (N=13), finding that happy routes increase subjectively perceived valence by 11\% (p=.007). Although happy routes take 1.25 times longer on average, participants perceived the happy route as shorter, presenting an emotion-enhanced alternative to today's fastest routing mechanisms. We discuss how emotion-based routing can be integrated into navigation apps, promoting emotional well-being for mobility use.

% Routes represent an integral part of triggering emotions in drivers. Many navigation systems allow users to choose a navigation strategy, such as the fastest or shortest route. However, they do not consider the driver's emotional well-being. We present \system{}, a novel navigation-based empathic car interface guiding drivers through real-world traffic while evoking positive emotions. We propose a set of design considerations, derive a technical architecture, and implement an optimization framework. Our contribution is a machine learning-based generated emotion map layer, predicting emotions along a route based on static and dynamic contextual data. We developed \system{}, a real-time mobile navigation app to predict routes evoking positive emotions interactively. We evaluated \system{} in a real-world driving study ($N=13$), finding that happy routes increase subjectively perceived valence by $11\%$ ($p=.007$). \textcolor{black}{Furthermore, although happy routes take longer ($1.25$ min on average), most participants ($8/13$) perceived the happy route as shorter, presenting an emotion-enhanced alternative to today's fastest routing mechanisms. }
% Finally, we show how emotion-based routing can be integrated into common navigation apps, promoting emotional well-being for general mobility use.
\end{abstract}



%\ccsdesc[300]{Human-centered computing~Empirical studies in HCI}
\begin{teaserfigure}
\centering
\centering
    \includegraphics[width=1\textwidth]{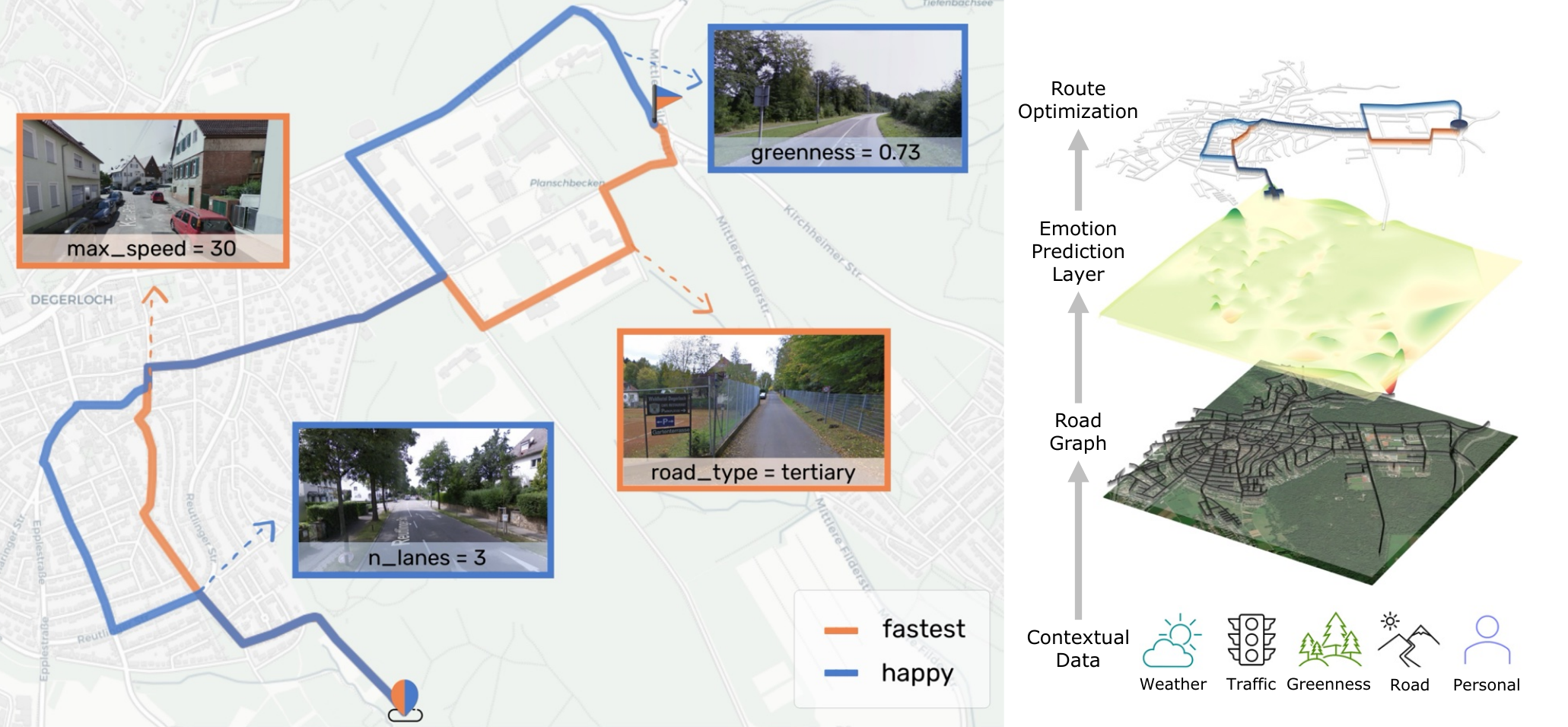}
\caption{We present \system{}, a new navigation system that routes after positive emotions. \chadded[id=R3]{The two fast and happy routes are exemplary and exhibit different environmental characteristics that can influence a driver's emotions.} We predict emotional weights for every road coordinate based on environmental, personal, and dynamic road context and find the optimal driving trajectory. 
}
\label{fig:teaser}
\Description[Teaser figure of the paper depicting the \system{} setup.]{Figure 1 illustrates AffectRoute and is entitled "Teaser Figure". The left side shows a map of a city where two routes are depicted (one fastest route and one happy route). In addition, pictures of road elements are shown for every road segment (e.g., number of lanes of the happy, maximum speed for the fastest route). The right side of the plot depicts the system's architecture in a 3d map layout. After inferring the road type from the depicted image on the left, a classifier predicts emotion weights based on context information. Lastly, the map shows a 3d custom emotion map layer where less happy segments are shown in red and high happiness assigned segments in green. An optimization then searches for the ideal "happiness" optimized driving route. }
\end{teaserfigure}

\maketitle

\section{Introduction}
\label{sec:intro}
Today's car navigation systems allow users to navigate according to various objectives, such as the fastest route, the shortest distance~\cite{golledge:pathselection}, or routes that require the lowest energy consumption~\cite{yang:evns}. In contrast to these routing modalities, we investigate a new objective by optimizing routes for positive emotions. Emotions play an important role in driving~\cite{hancock2012impact}, as certain positive or negative arousal and valence states can lead to more thoughtful decisions in the driving process, leading to safer driving. In contrast, exaggerated states such as anger can significantly increase the driver's willingness to take risks and thus endanger the safety of all road users~\cite{EHERENFREUNDHAGER2017236, pecher2009influence}. Those exaggerated states lead to more traffic accidents~\cite{underwood1999anger}. Subsequently, we propose \system{}, a system \chreplaced[id=1AC]{that navigates}{navigating} drivers through routes that elicit positive (i.e., happy) emotions to improve the driving experience and safety. \system{} refers to routing mechanisms that optimize for the user’s emotional well-being alongside conventional objectives such as travel time and distance.

While the vision, preferences, and design of empathic navigation have been presented in prior work~\cite{pfleging:experiencemaps}, its technical concept, implementation, and concrete evaluation have rarely been the subject of research. 
In particular, the field of in-vehicle emotion assessment~\cite{bethge_vemotion_2021,liu2021empathetic} has evolved strongly over the past decade, while empathic real-world applications remain the exception~\cite{zepf_driver_2020,braun2020if}. 
Based on an increasing number of available datasets that classify driver emotions based on driving context~\cite{bethge_vemotion_2021, balters:stresscar, liu2021empathetic}, we conceptualize and implement the missing building blocks for an end-to-end empathic navigation interface. 
Consequently, \system{} predicts possible emotions for thousands of unseen roads throughout a road graph and optimizes for the best tradeoff between positive emotions and travel time. While prior work has proposed emotion-focused navigation strategies based on static affective user ratings~\cite{doi:10.1080/13658816.2014.931585}, our system is the first to integrate real-time contextual data for emotion-driven route adaptation. \system{} dynamically updates routes based on predicted happiness levels using remotely acquirable data sources without requiring subjective user reports after model training. Thus, our approach dynamically incorporates contextual road features to predict emotions in real-time. This allows for personalized navigation that adapts without requiring manual user input during or after model training. The goal of  \system{} is to enhance emotional well-being by counteracting stress and frustration, known contributors to accident-prone behaviors~\cite{EHERENFREUNDHAGER2017236, pecher2009influence, underwood1999anger}.

\chadded[id=1AC]{Closest to our work is SAR by Wang et al. \cite{wang2018SAR}, a route recommendation system that considers social and environmental context factors affecting human emotions. The authors devised a heuristic model incorporating context information to determine the most enjoyable route. They evaluated their method based on computational performance and with five participants in a driving simulator. To improve and extend this idea, \system{} incorporates a machine learning model trained and quantitatively evaluated on a large dataset. In contrast to a driving simulator, we conducted a real-world user driving study to investigate how \system{} affects emotions while driving.} \chadded[id=2AC]{} \chadded[id=R3]{}

This paper presents design considerations, the resulting architecture, and an experienceable implementation for driving with positive emotions in real-world environments. We begin by discovering \chreplaced[id=1AC]{design considerations for}{the degrees of freedom to design} a scalable affective navigation system applicable to unknown users, environments, and roads. We demonstrate that theoretical psychological assumptions hold for the experienceable system, showing for the first time a navigation system that regulates emotions positively \chadded[id=R2]{by the choice of an optimized route}. Based on this, we derive the technical architecture for \system{}. An in-the-wild driving study with 13 participants investigates the effect on arousal and valence between choosing the \textit{fastest} route and the predicted \textit{happier} route. Our results show a significant effect in perceived valence between the fast and happy route, showing that the happy route selected by \system{} improves valence. Furthermore, our participants were willing to use \system{} although positive routes consumed more time. Moreover, we conducted a simulation study in a whole region to compare the differences between the optimization objectives. Finally, we conclude our work by discussing ethics, the applicability of \system{} for other transport modalities, generalization for unseen roads, limitations, and future work.

\begin{table*}[h]
\centering
\caption{\chadded[id=1AC]{Description of papers and their contribution to route recommendation after considering emotions. The dimension \textit{Model Applicable to Unseen Environments} refers to the fact that the developed model, either heuristic or as a classifier, does not rely on crowd-sourced user data (e.g., POI ratings).}\chadded[id=2AC]{} \chadded[id=R2]{} \chadded[id=R3]{}}
\resizebox{\textwidth}{!}{%
\begin{tabular}{@{}lccccccc@{}}
\toprule
\textbf{Paper} & \textbf{Emotion} & \textbf{Model Applicable to} & \textbf{Routing} & \textbf{Routing} & \textbf{Routing Quantitative} & \textbf{Routing Simulator} & \textbf{Routing In-the-Wild} \\
& \textbf{Classifier} & \textbf{Unseen Environments} & \textbf{Idea} & \textbf{Algorithm} & \textbf{Simulation Study} & \textbf{User Study} & \textbf{User Study} \\ \midrule
Bethge et al. \cite{bethge_vemotion_2021, bethge2022designspace} & \ding{52} & \ding{52}  & \ding{54} & \ding{54} & \ding{54} & \ding{54} & \ding{54} \\ 
\citet{doi:10.1080/13658816.2014.931585} & \ding{54} & \ding{54} & \ding{52} & \ding{52} & \ding{54} & \ding{52} & \ding{54} \\
\citet{liu2021empathetic} & \ding{52} & \ding{52} & \ding{52} & \ding{54} & \ding{54} & \ding{54} & \ding{54} \\
\citet{pfleging:experiencemaps} & \ding{54} & \ding{54} & \ding{52} & \ding{54} & \ding{54} & \ding{54} & \ding{54} \\
\citet{wang2018SAR} & \ding{54} & \ding{52} & \ding{52} & \ding{52} & \ding{54} & \ding{52} & \ding{54} \\
\textbf{HappyRouting (Ours)} & \ding{52} & \ding{52} & \ding{52} & \ding{52} & \ding{52} & \ding{54} & \ding{52} \\ \bottomrule
\end{tabular}%
}
\label{tab:prior_work_diff}
\end{table*}

\section*{Contribution Statement}

The contribution of this work is threefold: 
\begin{enumerate}
    \item We present \chreplaced[]{a set of comprehensive design considerations for}{the degrees of freedom to design} a scalable affective navigation system that applies to previously unknown users and unseen environments.
    \item With \system, we demonstrate that guiding design decisions hold for an experienceable end-to-end system and show for the first time that navigation systems can regulate emotions positively \chadded[id=R2]{by the choice of routes} \chadded[id=2AC]{in a real-world setting}\chadded[id=1AC]{}.
    \item We characterize the qualitative and quantitative properties of our proposed affective navigation system in an in-the-wild user study ($N=13$) and with detailed simulations.
\end{enumerate}

\section{Related Work}
\label{sec:related_work}
%Previous research investigated how emotions can be sensed to improve the driver's  experience. This section summarizes research in driver emotion assessments and eludicates how these insights contributed to the realization of empathic car interfaces.

% \todo[inline]{@Kosch: Table mit contribution type: data-driven model / real-world analysis / simulator study / design considerations}

\system{}'s idea of routing after positive emotions build on concepts found in driver emotion assessment, contextual computing, and empathic car interfaces. 

\subsection{Inferring Driver Emotions}
%Sensing driver emotions has moved into the focus of past research. 
Using context-aware sensing~\cite{schmidt2000implicit} using human sensing in cars~\cite{10.1145/3534617} gained increased attention to understanding driving behavior or facilitating novel perceptual car interfaces. In this context, empathic car interfaces benefit from understanding the driver's emotions to adapt their interface, contributing positively to the user's emotional state~\cite{10.1007/11573548_125, zepf_driver_2020}. Emotion assessment can be achieved through \textit{direct} and \textit{indirect} user observation.

\textit{Direct} observation methods, such as recognizing facial expressions~\cite{ekman1984expression,ekman1997face}, are a convenient method to infer emotions while driving. 
Although facial expressions are a commonly used modality~\cite{Braun2020}, it remains controversial in research~\cite{mollahosseini2017affectnet, heaven2020faces}.
Alternatively, emotions can be derived from psychophysiological signals such as electrodermal activity, heart rate, muscle tension, respiratory rate, and electroencephalography~\cite{s18072074,balters:stresscar, 10.1145/3379157.3391655}. 
The setup of in-car physiological sensing is often problematic due to insufficient signal quality levels~\cite{EGGER201935} and missing user acceptance~\cite{YANG2016256}. 

%can be a laborious endeavor. Especially for in-car settings, wearable devices must provide a sufficient utility to the user to justify the user's effort of using the wearable sensor~\cite{YANG2016256}. 
%Due to these circumstances, current automotive applications prefer remote-sensing technologies through camera or voice. 
%However they inhibit privacy concerns and achieve suboptimal results in a driving setting~\cite{stark2019facial, bethge_vemotion_2021}.

\textit{Indirect} user observation through analyzing contextual driving data \chdeleted[id=1AC]{has} gained increasing attention for emotion recognition. 
Zepf et al.~\cite{zepf2019towards} surveyed affective automotive user interfaces and identified several factors causing emotional triggers and changes, including driving behavior, music, and road conditions. 
This fact was exploited by Liu et al. by analyzing vehicle CAN-bus data~\cite{liu2021empathetic}, reaching subject-independent F1-scores of 59\%. 
Bethge et al.~\cite{bethge_vemotion_2021, 10.1145/3544549.3585672} showed that contextual driving data captured with a smartphone resulted in subject-independent F1-scores of 56\%, an improvement over using facial expressions as a baseline. In contrast to previous work~\cite{liu2021empathetic}, the authors utilized data from a smartphone only.

%An empathic navigation system poses additional constraints on the observation method since it is required to predict emotions on thousands of possible road segments to find the optimal emotion-aware route. Therefore, algorithms trained on direct observations as input data are not applicable, and only those operating on indirect observations gained through static (e.g., road properties) or live public data (e.g., traffic) can be applied.

An empathic navigation system poses additional constraints on the observation method since it is required to predict emotions on thousands of possible road segments to find the optimal emotion-aware route.
Since, in most cases, there is no direct observation input (e.g., crowd-sourced facial expressions) accessible for every unseen road segment, algorithms trained on remotely accessible observations (e.g., traffic, road properties, or weather) are needed. 

\subsection{Affective Routing}

Routing is considered a factor that strongly influences the driver's emotions. 
In their detailed study, Braun et al.~\cite{braun2020if} explored 20 concepts for empathic car interfaces, finding that empathic navigation is desired among German and Chinese drivers. 
In a web survey, Pfleging et al.~\cite{pfleging:experiencemaps} evaluated the general idea of experience-based navigation and identified the fastest route and the route with the least stress as the most important factors for route selection. 
At the same time, users often bypass the fastest route, for example, to avoid stressful situations and negative emotions~\cite{ceikute2013routing}. 
Zepf et al.~\cite{zepf2019towards} showed that most positive emotional triggers are associated with the environment. 
Accordingly, positive and negative experiences with a route play a crucial role for the acceptance of future route recommendations \cite{samson_recommendation_deviation}. 
Previous work \chdeleted[id=1AC]{has} focused on various routing concepts that may indirectly influence emotion. This stands in contrast to \system, which directly optimizes for positive emotions by applying a diverse set of features. 

Quercia et al.~\cite{quercia:theshortest} \chreplaced[id=1AC]{investigated}{investigate} a scenic routing concept using crowd-sourced images associated with POIs. Similarly, Runge et al.~\cite{runge_scenic_routing_google_maps} \chreplaced[id=1AC]{identified}{identify} scenic rides by applying a pre-trained neural network to street view imagery. \textcolor{black}{The routing methodologies~\cite{quercia:theshortest, runge_scenic_routing_google_maps} incorporate contextual routing concepts but are limited by crowd-sourced data and the visual attributes of the place itself.} 
\chadded[id=1AC]{Wang et al.\,\cite{wang2018SAR} presented a route recommendation system that optimises routes based on their social and emotional impact. The authors presented a heuristic model that determines a route based on factors such as traffic and historical emotions of a previously driven route. The authors found that 4 out of 5 participants who participated in a simulator driving study preferred the route suggested by the algorithm to a randomly assigned route.}\chadded[id=2AC]{}

\chadded[id=1AC]{\citet{doi:10.1080/13658816.2014.931585} presented a mobile application that uses an affect-space-model for collecting emotional responses - the basis for a route planning algorithm. A user study revealed that the generated routes are preferred over conventional shortest routes used in navigation systems.} \chadded[id=2AC]{}

Using physiological data, Tavakoli et al.~\cite{tavakoli:leveraging} \chreplaced[id=1AC]{introduced}{introduce} a framework for routing recommendations based on the driver's heart rate collected in a three-month in-the-wild study. 
The authors note that the proposed framework can find infrastructural elements in a route that can potentially affect a driver's well-being. 
Hernandez et al.~\cite{hernandez:autoemotive} proposed the long-term vision of crowd-sourced driver stress detection~\cite{nass2005improving} using ``Empathetic GPS'' - a vision of a navigation system that geographically identifies routes \chreplaced[id=1AC]{which minimize driver stress.}{minimizing stress whilst taking the driver to a given destination.}

\subsection{Summary}

Previous works show that empathic navigation is a highly desired feature among drivers and co-drivers~\cite{braun2020if,pfleging:experiencemaps} \chadded[id=1AC]{(see \autoref{tab:prior_work_diff} for a summary). } \chadded[id=2AC]{}Our work leverages such initial concepts and contributes with the technical building blocks to ultimately present \system{}, a real-world, end-to-end affective navigation system. To our knowledge, \system{} is the first experienceable system that predicts emotions for real-time routing in-the-wild.

\section{Design Considerations}
\label{sec:design}
\label{sec:concept}

% \todo{What are the user requirements for navigation? Do user want scenic routing? Bastian did this, clarify this here -> Moved to summary in Section 2.3}

The following section describes our design considerations for \system. We start by describing how \system{} will affect the driver's emotions, mood, and well-being. Then, we look into different routing concepts and conclude with relevant objectives and the modeling of driver routes. A particular focus on ethics and limitations can be found in our discussion in Section \ref{sec:discussion}. 

\subsection{User Emotions, Mood \& Wellbeing}
Our goal \chreplaced[id=1AC]{was}{is} to create a joyful driving experience that is implicitly composed of contextual data such as traffic, road characteristics, and weather. 
In general, our approach can be considered a method for regulating emotions\,\cite{mcrae2020:emotionegulation} during navigation, particularly aiming for an upregulation. 
Emotions can be regarded as situationally bound, limited in time, with either a positive or a negative state\,\cite{mcrae2020:emotionegulation}. Hence, emotions can change throughout a ride.
For example, the traffic flow or route characteristics, which are among the primary sources of information for \system, can manipulate perceived driver emotions.

In contrast to emotions, the user's mood is less intense and specific and often not caused by a particular event or situation\,\cite{frijda1993moods}, such as the current weather\,\cite{keller:moodweather}. 
\system{} primarily aims to elicit positive emotions, eventually positively influencing the user's mood. 
However, this approach is deliberately oversimplified as it is necessary to consider the overall process of mood adjustment and counter-hedonistic effects\,\cite{knobloch:moodadjustment}.
This means that positive mood is not only established by a simple aggregation of positive emotions but rather a complex interplay of positive \textit{and} negative emotions (e.g., people like to listen to sad music to adjust their mood positively)\,\cite{nabi:emotionalflow}.

Another constraint of our approach is focusing on primary emotions, particularly positive affect. However, positive affect and the absence of negative affect represent only a subset of possible dimensions to improve subjective well-being \,\cite{gallagher_wellbeing}. Other important factors, particularly in the dimensions of social well-being and eudaimonic well-being (e.g., self-acceptance), are currently well outside our scope of work. In summary, \system\, can be seen as a first important step towards a more detailed understanding of how technical systems can positively impact emotions. On the other hand, the aforementioned limitations raise many important questions for future work.

\subsection{Routing Concepts}
All routing concepts have in common that they operate on a graph of nodes and edges with associated weights. 
Edges represent a road segment in the road network, while nodes connect the various road segments. 
The weights associated with a road segment can represent different optimization objectives, for example, routes with the fine particulate (PM2.5) intake~\cite{mahajan:car} or those requiring the least energy~\cite{yang:evns}. 
For most use cases, the primary optimization objective is tightly coupled with travel time and distance between the two nodes. 
Eventually, this fact leads to the need for multi-objective optimizations, achievable through single- and multi-stage optimization. 

\textit{Single-stage optimization} combines multiple optimization objectives into one optimization method. 
For this purpose, multiple weights corresponding to the different objectives are associated with each edge, sometimes called layers. 
In the simplest case, the final weight can be determined by weighted addition of the individual weights \cite{voelkel:routecheckr} or introducing a penalty factor \cite{kaparias:timedependence}. 
If multiple objectives have statistical dependencies, more complex models like Bayesian Belief Networks can determine the combined weight~\cite{sharker:healthoptimalrouting}. 

\textit{Multi-Stage Optimization} conducts multiple optimizations in succeeding steps, with the first steps representing the most important optimization objectives. 
This optimization procedure can be used if the optimization problem is expressed through multiple models, e.g., road graphs and lists of POIs. 
For example, Quercia et al. \cite{quercia:theshortest} apply Eppstein's algorithm \cite{eppstein:shortest} to find the $N$ shortest paths, and then, in a second stage, rank those paths by user scores for POIs. \chadded[id=2AC]{A modified approach for the $N$ shortest paths problem was presented in SAR\,\cite{wang2018SAR}.}

For \system, we apply single-stage optimization, as we can associate emotions to each road segment, enabling us to express the problem in a uniform way. 
\system's primary objective is \textit{travel time}, while emotions are added to the graph's weights as a penalty term\,\cite{kaparias:timedependence}. 
The penalty term is computed through a machine-learning model, which considers various emotion-related features. 
Routes can be computed with efficient graph-based algorithms like Dijkstra or A*, or in our case, the contraction hierarchies algorithm specifically designed for vehicle navigation optimization\,\cite{geisberger2008contraction}.
%we can make it shorter: Routes can then be computed with efficient graph-based algorithms like contraction hierarchies algorithm specifically designed for vehicle navigation optimization

\subsection{Optimization Objectives}

Considering human wayfinding, Golledge \cite{golledge:pathselection} ranked various route selection criteria. 
\textit{Shortest distance} ranked first and \textit{least time} second, followed by \textit{fewest turns} and \textit{most scenic}. 
Less generic approaches consider criteria like \textit{least energy} \cite{yang:evns}, \textit{least fine particulate (PM2.5) intake} \cite{mahajan:car}, \textit{optimal physical exercise} \cite{sharker:healthoptimalrouting}, or \textit{personalized accessibility metrics} \cite{voelkel:routecheckr,kasemsuppakorn:personalizedwheelchair}. 
We can categorize these criteria into the following optimization objectives: 

\begin{itemize}
    \item \textit{Environment-dependent} objectives, e.g., \textit{shortest distance} do not change over the duration of the trip. In \system, we utilize properties of the environment like the number of lanes and speed limits to derive emotions. 
    
    \item \textit{Time-dependent} objectives change over the duration of the trip, such as \textit{least time} would be affected by time-dependent traffic. Our primary optimization objective in \system\, is the travel time, which is penalized by negative or neutral emotions. 
    
    \item \textit{User-dependent} objectives depend on personal criteria, such as accessibility needs. \system{} attempts to scale across various users, including unknown ones. Therefore, we include some user-dependent features, i.e., personal context as input in our architecture but identify a need for further exploration in future work. 
    
    \item \textit{State-dependent} objectives consider the state of an object, such as an electric vehicle's charging state \cite{yang:evns}. We do not consider this in our objectives for practical reasons and generalization purposes. %previous emotional states in our objectives. 
    %\item \textit{State-dependent} objectives consider the state of a person or an object, such as an electric vehicle's charging state \cite{yang:evns}. For practical reasons and generalization purposes, we do not consider this in our objectives. 
\end{itemize}

\begin{figure*}[t]
    \centering
    \includegraphics[width = 0.75\linewidth]{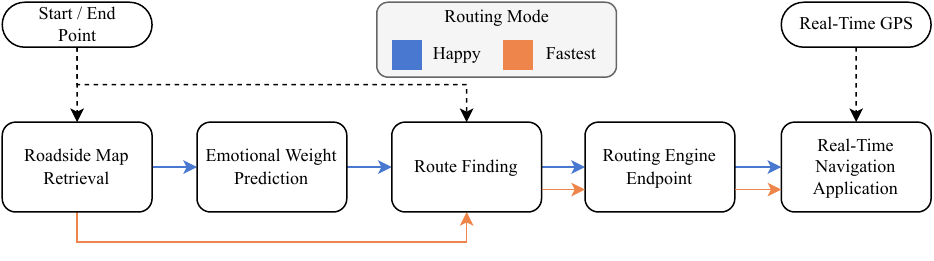}
    \caption{Architecture of happy navigation computation.}
    \Description[Graphical representation of \system.]{Figure 2 shows a graphical description of the happy navigation computation. Between a start and \chreplaced[id=1AC]{end point}{endpoint}, the maps are retrieved to make an emotional weight prediction to find a happy route in case the happy route is desired. This step is skipped if the fastest route should be found. Finally, the routing engine \chreplaced[id=1AC]{end points}{endpoints} are computed to provide the corresponding navigation.}
    \label{fig:architecture}
\end{figure*}

\chreplaced[id=R2]{
Different optimization objectives raise the question of their societal impact, particularly when applied at a large scale. Johnson et al.~\cite{johnson_externalitites} discuss the potentially negative influence of scenic routing algorithms and their optimization objectives on neighborhoods and parks, such as increased traffic presence in otherwise quiet residential areas. Besides this, emotion-based navigation can also have an environmental impact, e.g., if routes are longer. We refer to Section \ref{sec:discussion} for a more detailed discussion, including the implications of automated driving systems.
}{Different optimization objectives raise the question of their societal impact, particularly when applied at a large scale. 
Johnson et al. discuss such potentially negative influence of scenic routing algorithms and their optimization objectives on neighborhoods and parks \cite{johnson_externalitites}. 
We refer to Section \ref{sec:discussion} for a more detailed discussion.}

\subsection{Modeling and Simulation}

Most sophisticated optimization objectives require an approach to express their influence on the weights of a graph. 

The most common form is modeling based on historical observations, especially of travel times \cite{pfoser:dynamictraveltime} or least fine particulate (PM2.5) intake \cite{mahajan:car}. 
However, the two examples differ greatly in how they can be applied to a graph network. 
Regarding travel times, observations can directly be linked to edges in the graph. 
For fine particulate (PM2.5), an intermediate interpolation and edge association step is needed, as observations are linked to measurement stations \cite{mahajan:car}. 
\system\, applies both methods for different features: On the one hand, the characteristics of the road segments are used as direct parameters, and on the other hand, metrics of the surrounding landscape, such as the green index, are interpolated. 

Models that take travel time into account require time-dependent modeling, as traffic and, therefore, weights in the graph change over the duration of the trip. 
Such look-ahead models are often based on historical observations. 
Except for \textit{travel time}, \system{} currently does not consider additional fast-changing environmental parameters. 
In particular, the weather will be considered static throughout the trip. 
This design choice reflects a lowered computational effort at the cost of potentially less accurate predictions in the future. 

The penalty term for each road segment can be represented as a regression model in many use cases like travel time prediction \cite{pfoser:dynamictraveltime}. 
In contrast, \system\, is based on a multi-class model for predicting emotions (e.g., happy, neutral, sad), where the inputs consist of road parameters and the outputs represent the pseudo-likelihood of each class. 
To synthesize the penalty term for the graph edges, we used only the pseudo-likelihood of the class \textit{happy}. 
Alternative methods represent the model as a binary classifier (e.g., happy against all other classes) at the cost of decreased performance due to an increased imbalance of classes.

\section{Architecture}
\label{sec:architecture}
In the following section, we describe the architecture of our system and the necessary steps to provide the user with an emotion-optimized route. 
We derive the technical considerations from the concept design considerations presented in Section~\ref{sec:concept}.

\subsection{\chreplaced[id=R3]{Requirements}{Problem Statement}}
Finding an \chreplaced[id=1AC]{emotionally-relevant}{emotional-relevant} route is complex due to several reasons. 
The route optimization must be executed in near real-time, and all information required for the routing algorithm must be available (see Section~\ref{sec:concept}). Given a user's starting point $a$ and selected destination $b$ as GPS coordinates, we search for a route that likely makes the user happy. The following requirements must be fulfilled by \system: 
\begin{enumerate}
    \item[\textbf{Req 1:}] The emotional component of the route is subject to context, person, and traffic characteristics
    \item[\textbf{Req 2:}] The happiness weight of road segments has to be assigned before starting the navigation 
    \item[\textbf{Req 3:}] The system should be usable like a common smartphone navigation system 
    \begin{enumerate}
        \item The system should enable destination search functionality (e.g., finding a train station)
        \item The system should re-locate given the \chreplaced[id=1AC]{smartphone's}{smartphones} geolocation and show the trajectory of the happy route 
        \item The system should output turn-by-turn navigation instructions to the driver in real-time
    \end{enumerate}
    \item[\textbf{Req 4:}] The navigation engine should be designed as a scalable system 
    \begin{enumerate}
        \item Provide happy routes in every geolocation (no pre-annotated or historic routes)
        \item Optimize the route trajectories without delay so that the user receives the route recommendation $< 2 s$ after entering the destination
    \end{enumerate}
\end{enumerate}

\subsection{General Framework}
\autoref{fig:architecture} provides an overview of the system architecture. 
Depending on the start to end point, a roadside map is created via OpenStreetMap (OSM)\footnote{\url{https://www.openstreetmap.org}}. 
We then perform a custom map layer computation in the subsequent step in which we predict emotional weights for every edge in the road graph. 
The happy route is then found with the newly created map via an optimization procedure. 
We expose the endpoint of the navigation engine and build a real-time navigation smartphone app on the basis of the routing engine.

\begin{table*}[t]
\centering
\caption{List of available features to predict drivers' emotions.}
\label{tab:list_available_features}
\Description[Tabular description of the input features in \system's emotion prediction model.]{Table 1 shows a tabular description of the input features in \ system's emotion prediction model. The table has several columns: context, feature name, example value, description, and source for all the features used in the system. The rows represent the different accessed variables, e.g., weather (feel temperature outside, wind speed, cloud coverage, and weather term), traffic features (free-flow speed and reduced speed), road features (road type, maximum speed, and the number of lanes), greenness score (satellite-derived) and personal context features (daytime, age, before emotion). }
\begin{tabular}{lllll}
\hline
\textbf{Context} & \textbf{Feature}          & \textbf{Example} & \textbf{Description}                                              & \textbf{Source}               \\ \hline
weather          & feeltemp\_outside         & 13.0                    & temperature outside of car                                        & Azure Weather      \\
                 & windspeed                 & 5.6                     & windspeed in km/h                                                 &                            \\
                 & cloud\_coverage           & 76                      & relative cloud coverage in \%          &                            \\
                 & weather\_term             & `clear'                 & description of weather condition                                  &                            \\ \hline
traffic          & reducedspeed & 7.295495                & current reduced speed to freeflow speed  & Azure Traffic      \\
                 & freeflow\_speed           & 115.0                   & freeflow speed expected under ideal conditions               &                            \\ \hline
road             & road\_type                & `residential'           & road type of current position                                     & OpenStreetMap             \\
                 & max\_speed                & 120.0                   & maximum allowed speed on the road                                 &                            \\
                 & n\_lanes                  & 2                       & number of available lanes                                         &                            \\ \hline
greeness         & satellite\_greeness       & 0.2                     & percentage of green pixels in \chadded[id=1AC]{the} environment & Mapbox Satellite \\ \hline
time        & daytime & `afternoon' & current daytime & system input \\
\hline
%personal         & daytime                   & `afternoon'             & current daytime                                                   & user input          \\

  personal               & age                       & 21                      & age of the driver                                                 &  user input                          \\
                 & before\_emotion           & `happiness'             & subjective expressed emotion before driving         &                            \\ \hline
\end{tabular}
\end{table*}

\subsection{Input Features}
\label{subsec:input_features}

We used a reduced number of contextual road features of the original dataset~\cite{bethge_vemotion_2021} for our custom context-emotion classifier model. 
%feature selection mechanism, %some features remain constant during the ride, why do we include them -> to be participant-dependent
\chreplaced[id=1AC]{The selected features were based on Braun et al. work~\cite{braun2021affective} where driving behavior, traffic, vehicle performance, and environmental factors were found to be discriminative of emotions.}{The selected features are based on Braun et al. work\cite{braun2021affective} where driving behavior, traffic, vehicle performance, and environmental factors are found to be discriminative of emotions.}
We filtered the variables based on the following requirements: (1) real-time, on-device computation without accessing the vehicle itself, (2) no direct user interaction, (3) non-critical consumption of device resources, and (4) time-critical computability. We \chreplaced[id=1AC]{restricted}{restrict} the model to only those input features that can be pre-computed before driving (\textbf{Req 2})\footnote{Contextual variables such as the current acceleration cannot be pre-computed.}. Furthermore,  personal factors such as age \chreplaced[id=1AC]{were}{are} used to adapt to user-dependent emotion-route preferences. The selected features are shown in 
~\autoref{tab:list_available_features}. We \chreplaced[id=1AC]{computed}{compute} the weather and traffic features for every road segment using Microsoft Azure's Weather and Maps API. Although the weather often stays similar across a larger geographical area, we \chreplaced[id=1AC]{included}{include} weather in the emotion prediction as the weather condition affects the route choice of our algorithm. For example, in rain, users may favor broad streets, while sunny conditions may prompt a preference for curvy, narrow roads. Thus, the emotion classifier learns this input interdependence.
\chadded[id=R3]{The feel temperature was provided directly by the Azure Weather API and comprises the levels of humidity, light, wind speed and real temperature.}
The road type features \chreplaced[id=1AC]{were}{are} gathered from OSM by selecting the nearest OSM node with its corresponding parameters. 
Based on satellite imagery, we \chreplaced[id=1AC]{quantified}{quantify} the vegetation and determine the green index~\cite{czekajlo2020urban} at any given geolocation.
\chadded[id=R3]{To obtain the greenness, we \chreplaced[id=1AC]{computed}{compute} the relative amount of pixels associated \chreplaced[id=1AC]{with vegetation}{to vegegation} in each given satellite image.}
We \chreplaced[id=1AC]{computed}{compute} the curviness using a weighted measure of the length of curves, which depends on the radius of a circumscribed circle that passes through all three consecutive geocoordinates in a route. Given $a,b,c$ as the length of the three sides of a triangle, the radius of the circumcircle is given by the formula:
\begin{equation}
    r = \frac{abc}{\sqrt{(a+b+c)(b+c-a)(c+a-b)(a+b-c)}}
    \label{eq:curvature_compute}
\end{equation}

%\todo[inline]{T: Add curviness prediction from the appendix}

\subsection{Emotion Prediction}

%\todo[inline]{T: Check if this subsection still makes sense after putting the feature paragraph on top}
 
The foundation for our emotional routing is a computational behavior model for predicting \chreplaced[id=1AC]{emotions}{emotion} using road context. We thereby learn subject-independent emotion labels for previously unseen road segments (\textbf{Req 4}). 
Recently, Bethge et al.~\cite{bethge_vemotion_2021} proposed an in-car remote-sensing system able to predict emotions on unknown roads for unknown users with very high confidence. 
The model is able to predict discrete emotion categories (`happy', `sad', `neutral', `angry', `contempt', `disgust', `fear', `surprise') using contextual road information (\textbf{Req 1}). Although many affective representation models exist (e.g., Plutchik's wheel of emotions describing 56 emotions~\cite{plutchik2013theories} or Russel's circumplex model~
\cite{russell_circumplex_1980}), we \chdeleted[id=1AC]{have} selected the seven emotion categories, as well as the category neutral. \textcolor{black}{Our model is designed to predict multiple emotions to ensure adaptability for navigation use-cases where other emotions predictions are needed, rather than simplifying it to a binary classification setting for just predicting `happiness'. For example, future research could use our classification model to build a navigation framework that avoids 'sad' emotional states for people with an anxiety disorder or emotional sensitivity~\footnote{We note that 'happiness' and 'sadness' are not antagonistic emotional states.}.}\chdeleted[]{.} The choice of our set of discrete emotions is practically grounded in Ekman’s theory which is often associated with emotion detection by analyzing facial features. We exploit this well-known model for our optimization and build a bridge to previous work~\cite{10.1007/978-3-030-17287-9_10, zepf_driver_2020}. 
\begin{table*}[bt]
\begin{center}
\caption{Mean (standard deviation) accuracy, class-weighted precision, recall, and $F_1$ scores of the cross-validation on unseen participants, i.e., leave-one-participant-out cross-validation. The model predicts eight emotion classes in total \textcolor{black}{ (left) or happiness vs. non-happiness emotions (right)}. \chadded[id=R3]{We applied a FERPlus-trained classifier\,\cite{ferplus} to the dataset, confirming findings in \cite{bethge_vemotion_2021}.}}
\begin{tabular}{lcccccccc}
\toprule
     & 
     \multicolumn{4}{P{3.8cm}}{\textbf{Leave-One-Participant-Out Cross-Validation (all emotion classes)}} & \multicolumn{4}{P{3.8cm}}{\textbf{Leave-One-Participant-Out Cross-Validation (happiness only)}}\\
    \textbf{Input} & Accuracy & Precision  & Recall & $F_1$ & Accuracy & Precision  & Recall & $F_1$  \\
    \midrule
    \makecell[l]{Facial Expr. \\ (FERPlus~\cite{ferplus})} & $.55 \pm .18$ & $\textbf{.53} \pm .19$   & $.55 \pm .18$   & $.49 \pm .19$  & $.20 \pm .15$ & $.20\pm .34$ & $.09 \pm .42$ & $.12 \pm .38$ \\
    
    %\makecell{\chadded[id=R3]{ Facial Expr.} \\ \chadded[id=R3]{(FERPlus~\cite{ferplus})}} & $.55 \pm .18$ & $\textbf{.53} \pm .19$   & $.55 \pm .18$   & $.49 \pm .19$  & $.20 \pm .15$ & $.20\pm .34$ & $.09 \pm .42$ & $.12 \pm .38$ \\
    \textbf{Our model}  & $\textbf{.63} \pm .16$ & $.49 \pm .21$ & $\textbf{.63} \pm .16$ & $\textbf{.53} \pm .20$ & $\textbf{.65} \pm .14$ & $\textbf{.22} \pm .20$ & $\textbf{.75} \pm .42$& $\textbf{.66} \pm .10$ \\
    \bottomrule
\end{tabular}
%}
\Description[Tabular description of the performances of the emotion classification system.]{Table 2 shows a tabular description of the input features in \system{}'s emotion prediction model. The columns represent the performance metrics accuracy, precision, recall, and F1 score in a leave-one-participant-out cross-validation. The rows include the metrics for facial expressions (baseline) and our emotion prediction model. Our model beats the baseline in all performance metrics but precision (0.53 vs. 0.49). The table includes the standard deviation of the performance metrics evaluated in the cross-validation setup.  }
\label{tbl:clf_report_evaluation}
\end{center}
\end{table*}
%\todo[inline]{T: Discuss dittrich\_validity\_emotion (now in bib) - they propose categorical emotion responses with 40 classes. Maybe one of the reviewers who was proposing these many classes was Sebastian Zepf or even Monique (who clearly had a conflict)}
In their in-the-wild driving study, the authors collected contextual driving data and subjective emotional states expressed by drivers while driving~\cite{bethge_vemotion_2021}. 
To not distract the driver and bias the ground-truth labeling, a beep tone every 60 seconds was triggered for the driver to verbally express their emotion according to a predefined set. 
We acquired the dataset and extended it by another 14 participants to 26 participants in total, reflecting in 31 sessions consisting of 438 min of emotion-labeled driving and eight classes of emotions in total\footnote{We note that the sole purpose of the training-dataset is to learn an emotion prediction engine that links contextual properties and emotion labels. Thus, this dataset differs from the in-the-wild experiment evaluation of \system{}.}. The dataset used in our study exhibits imbalanced labels due to the infrequent occurrence of negative emotional states in naturalistic driving environments. Specifically, our dataset includes approximately $120$ minutes of driving data labeled with 'happy' emotional states. \chadded[id=1AC]{We describe detailed information about the classifier dataset and the labeling procedure in the Section~\ref{subsec:appendix_classifier_dataset_description}.}\chadded[id=R3]{} After defining the input features, we selected a Random Forest Ensemble Learning as classifier based on a 10-fold grid-search cross-validation (using Support Vector Machines, Feedforward Neural Network, Decision Tree, Adaboost, and Random Forest classifier from scikit-learn with hyperparameter optimized parameters) in which the Random Forest achieved the highest average F1 score. \textcolor{black}{Furthermore, our Random Forest model is easily deployable on-device, making it an ideal choice for the real-time processing and low latency navigation application, without requiring specialized hardware on-device, unlike deep neural network architectures. } The prediction model~\footnote{Model parameter: class\_weight = 'balanced', max\_features= $log_2$, n\_estimators= $50$.} is tested via a leave-one-subject-out cross-validation on unseen participants. \textcolor{black}{Within one cross-validation fold a Random Forest model is trained on 25 participants and the performance is tested on the left-out participant's emotional data. Thereby, we prevent overfitting on individual participants' emotional data and assess how well the model generalizes, i.e., in predicting the emotions of unseen participants. This evaluation \chreplaced[id=1AC]{allowed}{allows} us to gauge the accuracy of a correct emotion prediction for a user that has yet to use the navigation system.} The results are outlined in \autoref{tbl:clf_report_evaluation}.
%Furthermore, compared to deep neural network architectures, our Random Forest model has the advantage of being relatively easily deployable on-device.
Overall, our model \chreplaced[id=1AC]{achieves}{is able to achieve} a mean emotion recognition accuracy of $63\%$ with a balanced $F_1$ score of $53\%$\footnote{Neutral emotions represent the majority class of our dataset, while happy emotions are at $23\%$,
being predicted second best (after neutral) in terms of precision and recall}. 
\chadded[id=2AC]{These results are slightly inferior to current subject-independent contextual emotion classifiers\,\cite{bethge_vemotion_2021, liu2021empathetic}, but are also based on a remotely acquirable, and thus much reduced, feature set.}
As a baseline in our dataset, we recorded a driver-facing camera stream and applied a FERPlus-trained classifier\,\cite{ferplus}, showing that the collected contextual features still outperform facial expressions\,\cite{bethge_vemotion_2021}. 
\begin{figure}[t]
    \centering
    \includegraphics[width=0.7\linewidth]{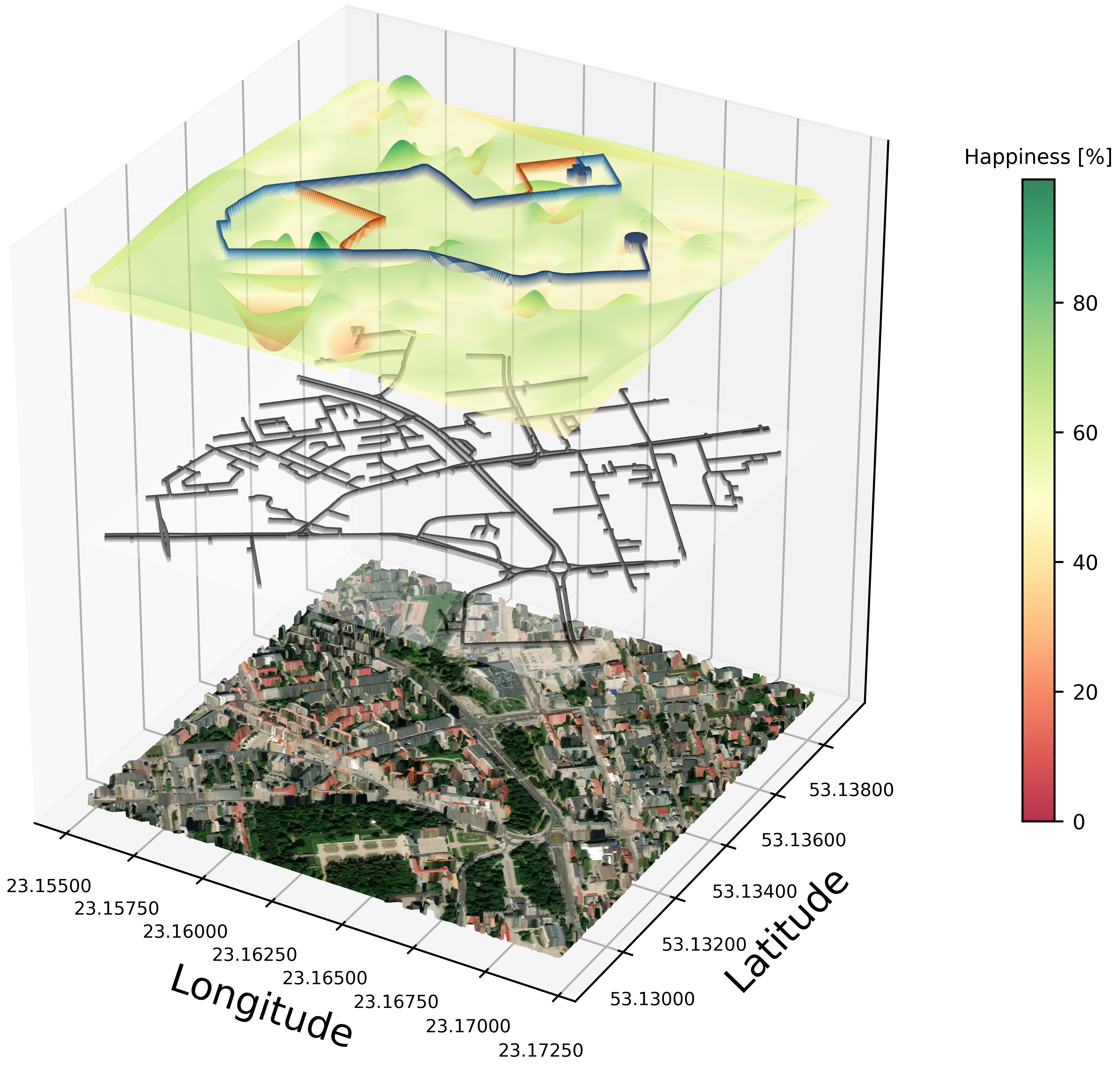}

    \caption{Graph Building for Happy Route Optimization. The navigation finds the optimal emotional path according to the emotion-road-weight regularization (\autoref{eq:routing_opt}). The bottom layer is a satellite image. The layer above represents the routable roads. Above is an emotion heatmap based on interpolation of the computed happiness points. The red path is the fastest path offered by navigation, while the blue path is the happy path. }
    \Description[Depiction of the AffectRoute graph building mechanism.]{Figure 3 shows three layers floating on top of each other. The bottom layer shows the satellite image of the map. The middle layer shows possible routable routes. Finally, the top layer shows a visualization of the routable roads.}
    \label{fig:emotion_stack_map}
\end{figure}
We also report the performance of a binary classification (happiness vs. non-happiness label prediction). 
%While the overall performance is lower due to imbalanced class distribution (a majority of non-happiness emotions), 
Here, the results of our binary classification model are superior to the facial expression engine. Our model achieves a mean result with an accuracy of $65\%$ and a $F_1$ score of $0.66$ vs. the facial expression engine with an accuracy of $20\%$ and a $F_1$ score of $0.12$. The results of the facial expression engine are vastly inferior as happy emotions are nuanced facial expressions that are hard to detect with a non-participant-trained computer vision model. Therefore, we argue that successful prediction of happiness on the road may require a more nuanced and multidimensional approach that considers a range of subjective and objective factors, including individual differences, social context, and environmental factors (as we do in our model). 
\chadded[id=1AC]{While the F1 score may seem not optimal, it's essential to consider that classification takes place over numerous road segments, ranging from hundreds to thousands, on a given route. Despite performance fluctuations for specific segments, the overall results will generally even out when applied to many instances. Our metrics align with findings in prior works, particularly when evaluated in a generalizable leave-one-subject-out setting, such as ours\,\cite{bethge_vemotion_2021,liu2021empathetic}.}\chadded[id=2AC]{}

\begin{figure*}[t]
\centering
%\begin{minipage}{.56\linewidth}
\begin{minipage}[b]{0.56\linewidth}
  %\vspace{-10em}
  \centering
  \includegraphics[width=1\textwidth]{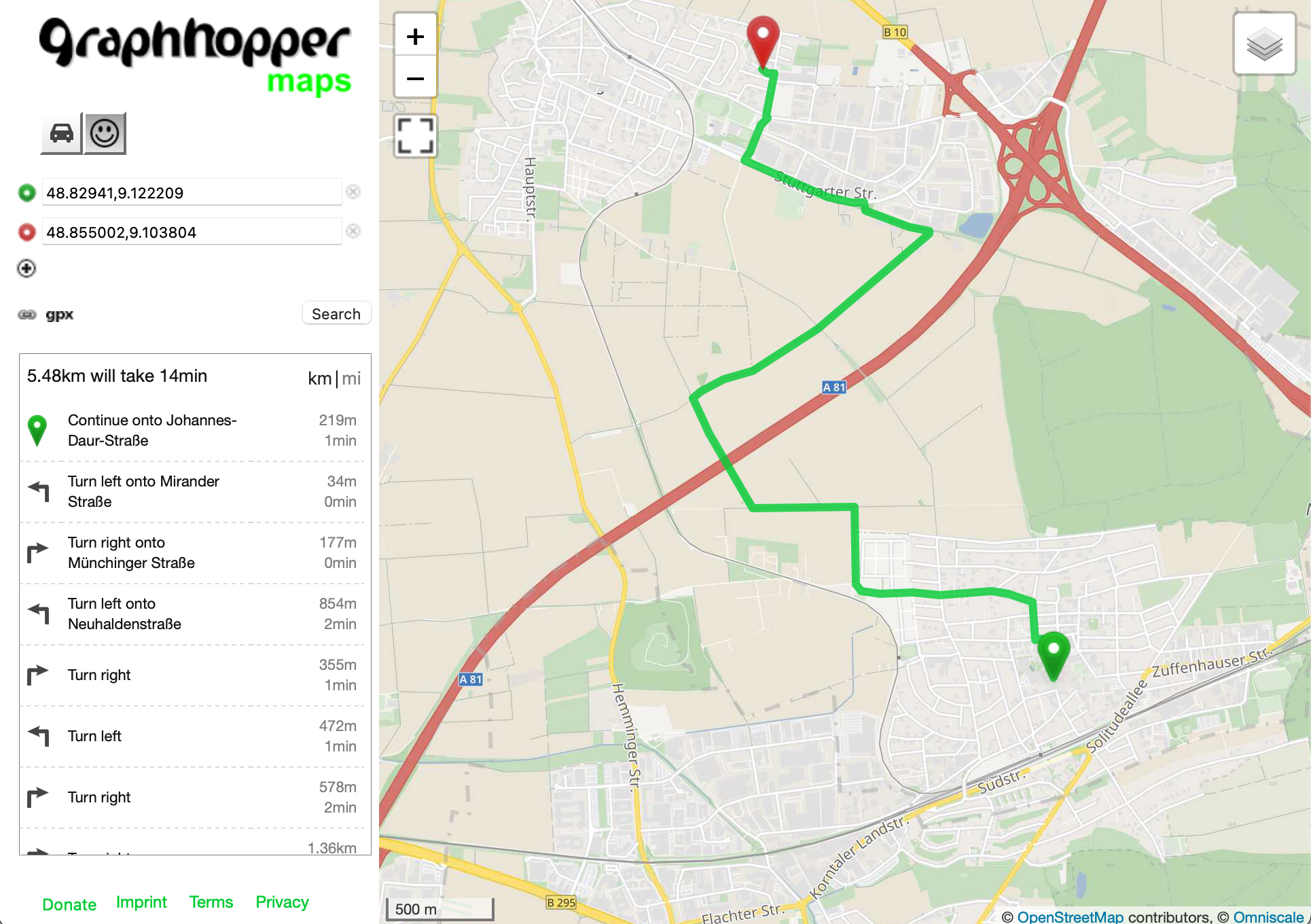}
  \captionof{figure}{GraphHopper web-server for Happy Route Optimization in a 2D-layout.}
  \label{fig:sub1_web_routing_engine}
  \Description[2D-layout of an optimized navigation route.]{Figure 4 shows a potential route of a happy route. The happy route is depicted with a green line.}
\end{minipage}%
\qquad
%\hspace{0.5cm}
\begin{minipage}[b]{0.395\linewidth}%0.4
  \centering
  \includegraphics[width=1\textwidth]{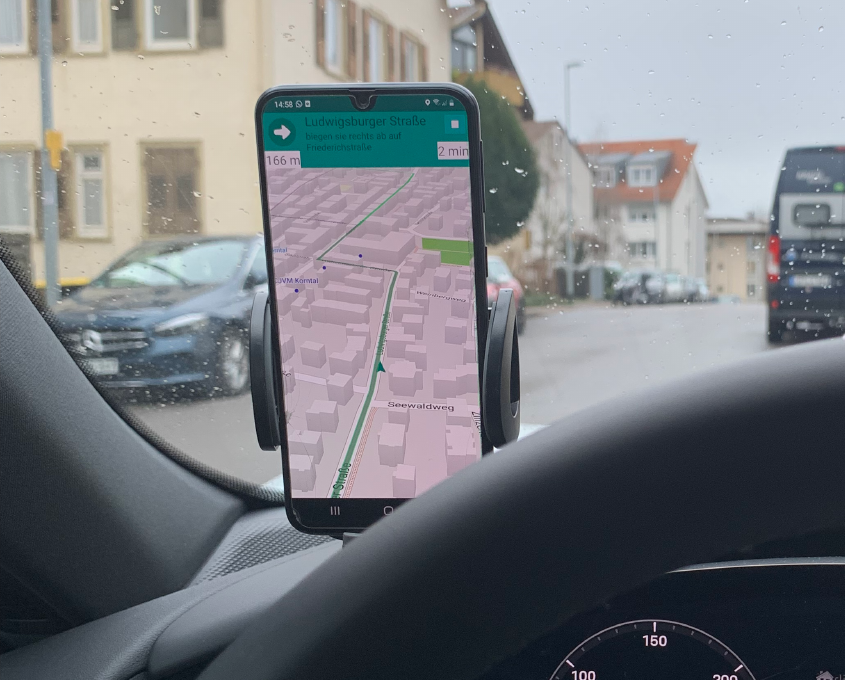}
  \captionof{figure}{Implemented navigation app that supports normal and happy routing. The app is placed on the windshield and has the same functionality as normal navigation apps (turn-by-turn navigation, voice output for hinting next directions).}
  \label{fig:phone_placement}
  \Description[Smartphone placed in a retainer while navigating through a route.]{Figure 5 shows a smartphone placed on the windshield of a car. The smartphone displays the AffectRoute through turn-by-turn navigation. A green line shows the route on AffectRoute.}
\end{minipage}
\end{figure*}

\subsection{Routing Map and Navigation}

Having defined the predictive model required to simulate emotions based on contextual information collected remotely, we now present the system required to provide users with a route optimized for emotions. 
In~\autoref{fig:emotion_stack_map}, we display how a happy path may differ from the fastest one based on a custom emotion map layer.

\paragraph{Routing Map Generation}

We \chreplaced[id=1AC]{defined}{define} a custom emotion map layer that contains predicted emotions and optimizes the route thereafter. 
Given a road graph $G$ with vertices $V$ and edges $E$, we \chreplaced[id=1AC]{predicted}{predict} emotion weights for every driveable segment $E$. 
We then \chreplaced[id=1AC]{applied}{apply} the contraction hierarchies algorithm\,\cite{geisberger2008contraction} to the road graph by optimizing for the following equation with the user's start point $a$ and endpoint $b$:

\begin{equation}
    route(a,b) = min \sum_{i, j \in [a,b]; i \neq j \in E }\frac{d(i,j)}{\lambda* e(i,j) *c(e(i,j))} %+ curvature + POI\_pred - %\frac{}{}
    \label{eq:routing_opt}
\end{equation}

In contrast to the fastest route, our optimizer minimizes the sum of the travel time of each edge $d(i,j)$ and penalizes its decision by the happiness weighing factor  $\lambda$ and its corresponding predicted happiness value $e(i,j)$, multiplied by the confidence of the individual happiness prediction $c(e(i,j))$.
Here, the last part ensures that it is favorable for the optimizer to choose edges with high predicted happiness values\footnote{We opt the routing decision formula to be influenced by the predicted emotion value in the denominator as the travel times have no equal lengths and regularizing longer route segments (high $d(i,j)$) with the emotion scaling is more beneficial than, e.g., subtracting the emotion values.}. 
In our simulation study (see Section \ref{sec:simulation}), we found that happiness weighing factor of $\lambda=20$ yields a good tradeoff between travel time and positive emotions.  

\begin{figure*}
    \centering
    \includegraphics[width = 0.8\linewidth]{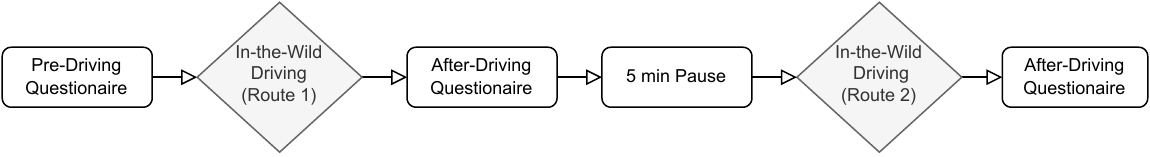}
    \caption{Experimental design of the emotional navigation driving study. The endpoint of the second drive was set to be the start point of the first drive.}
    \Description[Graphical representation of the experiment procedure.]{Figure 6 shows a graphical description of the experimental procedure. The first node entails the administration of the pre-driving questionnaires. The next node shows that the drivers are the first driving route. Afterward, they complete a questionnaire followed by a break of 5 minutes. The drivers then continue with the following route and complete an after-driving questionnaire.}
    \label{fig:experimental_study_design}
\end{figure*}

\paragraph{Optimization Backend}

To implement the optimization procedure, we \chreplaced[id=1AC]{used}{use} the open-source, Java-based framework GraphHopper\footnote{\url{https://github.com/graphhopper/graphhopper}}. 
GrapHopper offers a fast and memory-efficient routing engine, including a web frontend and a standalone web server to calculate a route's distance, time, turn-by-turn instructions, and trajectory properties. 
We \chreplaced[id=1AC]{adopted}{adopt} the routing optimization according to ~\autoref{eq:routing_opt}. 
We \chreplaced[id=1AC]{did}{do} not employ a standard A* algorithm~\cite{hart1968formal} for optimal route finding due to performance reasons. %for the optimal route finding, since it requires high space complexity as all nodes ad edges have to be stored. 
Instead, we \chreplaced[id=1AC]{disabled}{disable} all initial edge weight calculations for happy routing and \chreplaced[id=1AC]{built}{build} a prominently-used CH (Contraction Hierarchy) graph~\cite{geisberger2008contraction} with precalculated happiness weights to speed-up optimization (\textbf{Req 4}). 
\chreplaced[id=1AC]{We exposed}{Thereafter, our system exposes} a happy and fastest routing computation endpoint. 
The interactive GrapHopper routing endpoint for a happy route computation is shown in~\autoref{fig:sub1_web_routing_engine}. 

\paragraph{Smartphone Navigation App}
We implemented a scalable mobile application to provide users with the ability to navigate. 
Therefore, we customized the Android application  PocketMaps\footnote{\url{https://github.com/junjunguo/PocketMaps}} to use our optimization engine (\textbf{Req 3}). 
Our mobile application tracks the current smartphone geolocation using GPS and is able to search for destinations on the map via Google Maps search. 
The application then performs map matching of the current geocoordinate to the road segment and outputs turn-by-turn navigation instructions (via text and voice). 
Users can choose between the fastest and happiest routing in our app.~\autoref{fig:phone_placement} shows the navigation screen of our customized PocketMaps application in the wild.

\section{Driving Study}
\label{sec:in_the_wild_study}
The goal of our driving study is to gain an understanding of \system's user experience and its influence on a driver's emotions. We conducted a within-subject driving study to investigate differences in valence and arousal when using the fastest route compared to \system. 

\subsection{Participants}
Participants were recruited through a dedicated mailing list of colleagues willing to conduct research studies. Participants did not receive compensation for their involvement in the study. Participants gave their explicit consent to participate in the study and formally agreed by signing an informed consent form, which explained the details of the study and their rights. Participants were informed about the goals and procedure of the study. Participants could retract the study at any time. An independent review board granted ethical approval for the study, ensuring compliance with established ethical standards and protocols. We recruited 13 participants (11 self-identified as male, two self-identified as female) with an average age of $27\pm (8.51)$ years. Six participants drive occasionally (i.e., less than 10,000 km/year), six participants drive moderate distances (i.e., between 10,000 and 20,000 km/year), and one participant is a frequent driver (>20,000 km/year).
%The study took place on seven days.

\subsection{Methodology \& Procedure}
The participants accessed a vehicle with a standard Android smartphone attached to the windscreen (see~\autoref{fig:phone_placement}). 
We gave the participants time to get familiar with the car and explained that they could drive like they normally do (e.g., listening to music). 
We asked the participant to use our \system{} application just like a common mobile navigation app. 
\chadded[id=R3]{We selected start and end points that are approximately 15 minutes apart in terms of driving time, encompassing both rural and urban regions. This decision was influenced by the mean duration of journeys across the globe is approximately 15 minutes, although this value may fluctuate greatly depending on the country and other aspects~\cite{o2022and}. Considering that commuting accounts for most trips~\cite{eurostat2021passenger, ramos2020understanding}, we opted for an urban office location and a rural area as the two points in our real-world driving study.}

%\chadded[id=R3]{Globally, car usage and trip length heavily varies depending on country and other factors, however, the average trip duration worldwide is estimated to be approximately 15 minutes~\cite{o2022and}}.Therefore, the start and end location was chosen to be an approximately 15-minute drive away with segments including rural and urban segments. 
%\chadded[id=R3]{Commuting is the most common trip type~\cite{eurostat2021passenger, ramos2020understanding}, so that the start-point of our in-the-wild driving study is set to be at an office location. The trips include rural and urban segments with the endpoint of being in the countryside.}

The calculated routes were kept consistent for all participants to ensure comparability. The routing choice (fastest or happy routing) was hidden in the mobile application to avoid confirmation bias (i.e., blind route choice). 
The routing choice was randomized so that seven participants drove the happy route first, while six drivers were assigned to the fastest route first. \chadded[id=R3]{While the start and end points were the controlled variables in the trips, it is important to note that factors such as the time of day, the specific vehicle used, and resulting traffic conditions were not regulated within the study parameters. These deliberate choices were made to maintain the study's closeness to real-world driving scenarios, intentionally varying only the two route options while allowing other factors like time of day, the specific vehicle used, and traffic conditions to simulate natural, uncontrolled driving conditions.}

Overall, the one-way driving lasted approximately twelve minutes for the fastest route and 14 minutes for the happy route, depending on individual traffic conditions.
\chadded[id=R3]{Unlike the routes shown in~\autoref{fig:teaser}, the routes had very little overlap, with varying proportions of highways and secondary and tertiary roads.} 
The study protocol is presented in~\autoref{fig:experimental_study_design}. For each assessment of the driver's emotional state (valence, arousal), we \chreplaced[id=1AC]{applied}{apply} the self-assessment manikin (SAM) framework~\cite{bradley1994measuring} with a five-point Likert scale. \chadded[id=R3]{ More detailed questions, such as the participant's driving experience, trip-time estimate, or route favorability, were asked after the drive and can be found in~\autoref{tab:questionnaire}.}

% \begin{figure}[t]
%     \centering
%     \includegraphics[width=0.75\linewidth,trim=0 1cm 0 1cm,clip=true]{figures/experiment/phone_setup2.pdf}
%     \caption{Implemented navigation app that supports fast and happy routing. The app is placed on the windshield and has the same functionality as common navigation apps (turn-by-turn navigation, voice output for hinting next directions).}
%     \Description[Smartphone placed in a retainer while navigating through a route.]{Figure 6 shows a smartphone placed on the windshield of a car. The smartphone displays the AffectRoute through turn-by-turn navigation. A green line shows the route on AffectRoute.}
%     \label{fig:phone_placement}
% \end{figure}

\subsection{Results}

\paragraph{Valence-Arousal-Dominance Analysis}
 
\begin{figure*}[t]
    \centering
    \includegraphics[width = 1\linewidth]{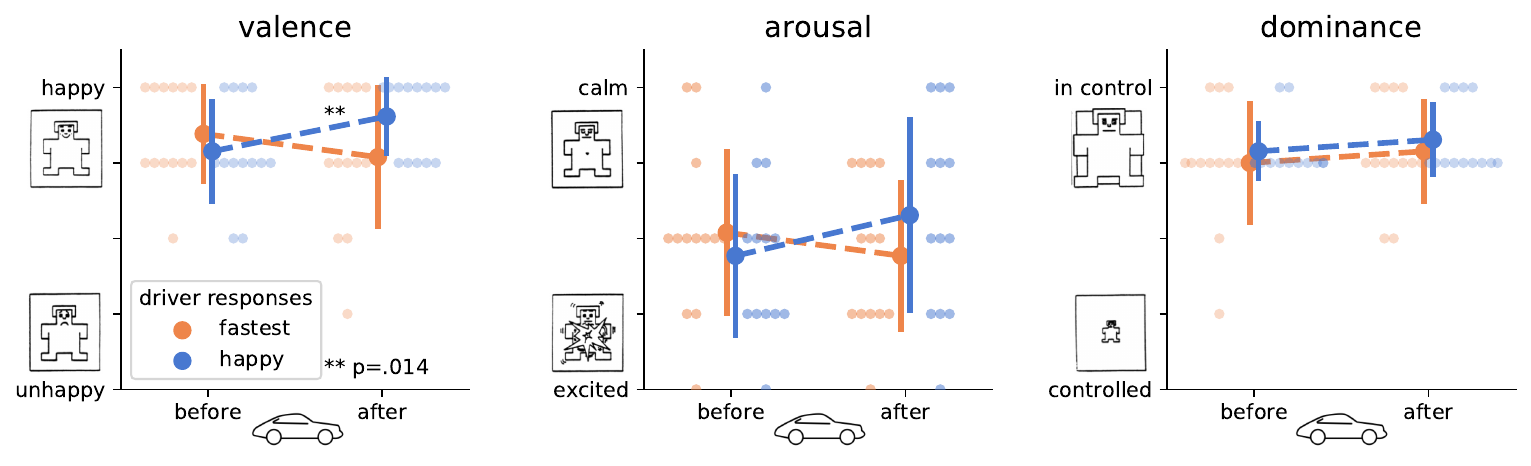}
    \caption{Before and after driving analysis of valence (left), arousal (middle), and dominance (right) questionnaire answers of the driving study. The lines indicate the responses' standard deviation (vertical) where the means are connected via the dashed line. The asterisk indicates significance. Fast and happy routes were assigned blindly and by random succession.}
    \Description[Three graphs showing the difference between valence, arousal, and dominance before and after using AffectRoute]{Three graphs show the difference between valence, arousal, and dominance before and after using AffectRoute. The left graph shows the difference in valence before and after driving the fastest or happy route. The graph denotes a significant effect on valence compared to the fastest and happiest route. The middle and right image shows the differences in arousal and dominance before and after driving the fastest and happiest routes. The changes in arousal and dominance are non-significant.}
    \label{fig:valence_arousal_before_after_in_the_wild}%The results do not show a significant effect.
\end{figure*}

We present the before and after analysis of valence, arousal, and dominance scores assessed with the self-assessment manikin questionnaire in~\autoref{fig:valence_arousal_before_after_in_the_wild}. 
We \chreplaced[id=1AC]{found}{find} that people gave higher valence ratings, i.e., positive attitudes, after taking the happy route. The mean valence score for happy routing before driving \chreplaced[id=1AC]{was}{is} $4.15$ and \chreplaced[id=1AC]{increased}{increases} to $4.62$ after driving ($11\%$ valence score increase). \chadded[id=1AC]{We statistically compared the driver emotions before and after driving to assess the impact of \system{} on driver emotions compared to fastest routing. We used a Shapiro-Wilk test for investigating deviations for normality.} \chadded[id=R2]{} Applying a Shapiro-Wilk test revealed a non-normal distribution \chadded{for our measurements}, $p < .001$.

\chadded[id=1AC]{We used Wilcoxon signed-rank tests for statistically comparing the emotion assessments within the routes. We calculate the effect size $r$ as suggested by Rosenthal et al.~\cite{rosenthal1994parametric}} \chadded[id=R2]{}.
A Wilcoxon signed rank test found a significant difference in valence before and after navigating through a happy route, \chadded[id=1AC]{$Z = -2.45$,} \chadded[id=R2]{} $p=.007$ \chadded[id=1AC]{, $r = 0.48$} \chadded[id=R2]{}. 
In addition, we did not find significant before-after differences in valence, \chadded[id=1AC]{$Z = -0.83$, $p = .405$, $r = .23$,} \chadded[id=R2]{} or arousal, \chadded[id=1AC]{$Z = -0.97$, $p = .334$, $r = .27$,} \chadded[id=R2]{} when driving the fastest route. Overall, we found a positive trend in arousal when driving the happy route, though all expressed arousal levels have high variance. The high variance likely \chreplaced[id=1AC]{results}{comes} from the fact that the driving task \chreplaced[id=1AC]{was}{is} perceived as relaxing or exciting on an individual driver's basis. Again, a Shapiro-Wilk test showed a non-normal distribution for arousal, $p=.025$. There was no significant difference in arousal before and after driving the happy route according to a Wilcoxon signed rank test, \chadded[id=1AC]{$Z = -1.65$, $p = .1$, $r = .46$} \chadded[id=R2]{}. This finding contrasts many empathic car applications that seek to optimize arousal levels for safety reasons~\cite{braun_improving_2019, braun2020if}.
\chreplaced[id=1AC]{We did not find significant before-after differences for the dominance scores when driving the fastest, $Z = -0.63$, $p = .527$, $r = .17$, and happy route, $Z = -1.41$, $p = .157$, $r = .39$.}{Furthermore, we did not detect any significant changes in dominance scores i.e. we detect no effect in how controlled or submissive one feels after driving the happy route.} \chadded[id=R2]{}

\chadded[id=1AC]{We statistically compare the perceived emotions for the fastest and happiest route after the driving trials.} \chadded[id=R2]{However, a Wilcoxon signed-rank test did not reveal a significant difference for valence, $Z = -1.89$, $p = .057$, $r = .52$, arousal, $Z = -1.81$, $p = .07$, $r = .5$, and dominance,  $Z = -1.00$, $p = .317$, $r = .27$.} \chadded[id=R3]{} 
\chadded[id=R2]{Route order had no statistical impact on participant ratings. In total, 54\% began with the happy route, showing a balanced setup.}

\paragraph{Happy Navigation Driving Behavior}
Our driving questionnaire showed high variability when and how drivers wanted to use happy navigation functionality. 
After the driving experiment, we asked the participants how much time they would sacrifice for a happy route, assuming $20$ minutes for the fastest route.
9 of 13 participants answered with $3$ to $5$ minutes, while 3 of 13 drivers would only spend $1$ to $3$ minutes additional drive time. 
One participant stated \chreplaced[id=1AC]{the willingness to}{that he would} even spend more than $10$ minutes of additional \chdeleted[id=1AC]{drive} time to drive the happy route. 
These results are consistent with the web survey by Pfleging et al.~\cite{pfleging:experiencemaps}, which states that participants would take \chadded[id=1AC]{on average} $20.9\%$ more time for an experience-optimized route compared to the fastest route.
% After riding both routes, we asked participants which route they found to be the longest ride. 
While the fastest route took on average 2 minutes less time, 8 of 13 participants perceived the happy route as shorter. 
Combined with the finding that subjects had a more positive emotional state after driving the happiness route, we conclude that a happiness route may positively influence the perceived travel time. 
Furthermore, in our study, 11 out of 13 participants stated that they would use the app in their leisure time when they did not have time pressure. 
Interestingly, many participants responded to use our navigation only on the weekend (P9, P10, P12), preferably in the summer (P1, P2, P3, P4, P9, P10, P12), and not at night when the driving scenery is not visible (P8, P13). 
\chreplaced[id=1AC]{P2 mentioned a preference for using happy routes in the event of a traffic jam, which would enable the choice of less crowded, more relaxed, and undiscovered routes.}{P2 mentioned that he would use happy routes ``if a traffic jam occurs and he could take lesser crowded, more relaxed and unknown routes''. }

\paragraph{System Acceptance}

In response to the question "How likely would you be to use this system?" on a scale of 1 (not at all likely) to 5 (very likely), 11 of 13 participants gave scores of 4 and 5. 
%Overall, we observed high demand for happy navigation, as only two participants gave a score of 3 and 2, respectively.
The study participants introduced ideas for pairing happy navigation with other \chdeleted[id=1AC]{further} in-car technology. 
\chreplaced[id=1AC]{The most prominent response was that many people associate happiness with music while driving.}{The most prominent responses were that many participants associate happiness with music while driving.}
Therefore, many suggestions were made to automatically select \chadded[id=1AC]{the} music to match the route, or vice versa, to select the route to match the music better.

We also asked the participants in a free-response question: "Do you think there are any societal and ethical implications of this navigation functionality? And if yes, which one?". 
Many participants said \chreplaced[id=1AC]{that they did}{they do} not see any ethical or societal implications (P6, P7, P9, P11, P13). 
Participants also responded with higher energy consumption costs and a more environmentally harmful behavior when driving a happy route (P1, P3, P10). 
\chreplaced[id=1AC]{P10 stated that there was a problem with happy routing only recommending pleasant routes so that other less happy predicted locations are not seen, creating a self-reinforcing effect of what people see.}{P10 stated to see a problem with happy routing only recommending pleasant driving routes so that other less happy predicted locations do not get visibility, creating a self-reinforcing effect of what people see.}

\section{Simulation Study}
\label{sec:simulation}
To offer a broad assessment of the recommended happy routes by our system, we \chreplaced[id=1AC]{performed}{perform} an offline numerical simulation analysis.

\subsection{Experiment Design} 
First, we \chreplaced[id=1AC]{downloaded and computed}{download and compute} the emotion prediction layer for a map of a medium-sized city ($12 \times 12$ km). 
We \chreplaced[id=1AC]{sampled}{sample} a large number of equally-distributed, random start and end points ($N=1000$) and \chreplaced[id=1AC]{searched}{search} for the happy and fastest routes. 
We then \chreplaced[id=1AC]{analyzed}{analyze} the route trajectories segments by computing several characteristics such as road types, greenness, traffic conditions, and curviness. 
Furthermore, we \chreplaced[id=1AC]{computed}{compute} the travel time, distance, and the overlap of the fastest and happy routes.

\subsection{Route Time Analysis}

\begin{figure*}[t]
\centering
\begin{minipage}{.47\textwidth}%.45\textwidth
    \centering
  \includegraphics[height=6cm]%{figures/results/byalstok_lambda=20_scatter_duration_divergence.pdf}
{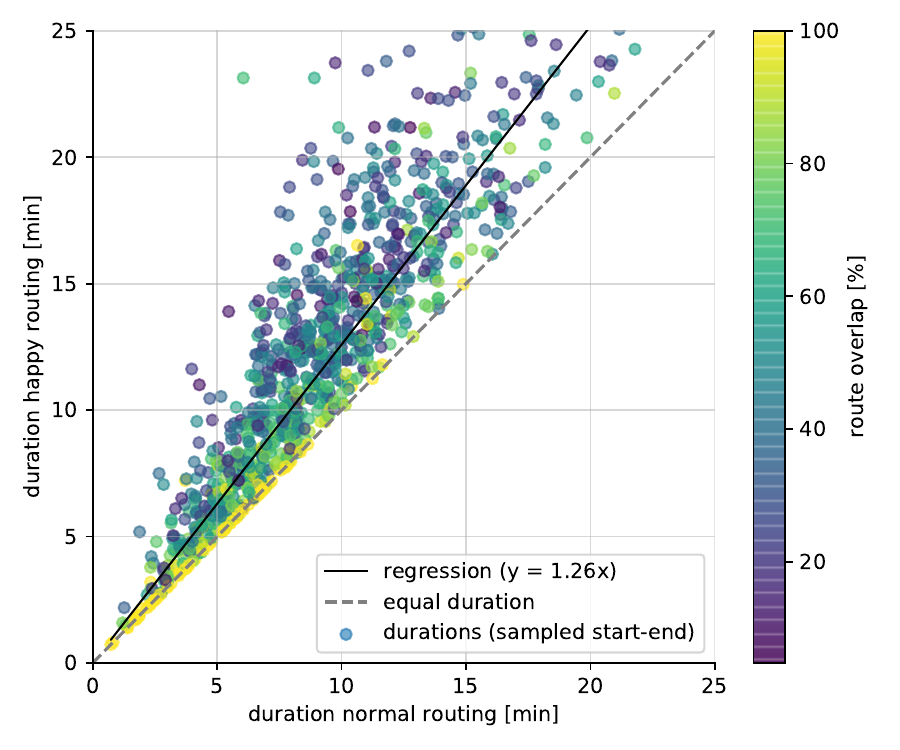}
  \caption{Scatter plot of drive duration of normal vs. happy routing. The points are mostly on the top-left of the equal-travel-time line, meaning happy routing generally takes longer to drive. We set $\lambda$ to $20$. The fitted regression ($R^2=0.81$, $BIC=1969$) with a slope of $\beta_1 = 1.26$ ($p=.00$) means that a 1-minute increase in normal routing will take $1.26$ minutes ($75.6$ $sec.$) more time to drive using happy routing.}
    \Description[Scatterplot of the drive duration between normal and happy routing.]{Figure 8 shows a scatterplot correlating the drive duration between normal and happy routing. The results show that driving one minute of a happier route will take 1.26 minutes when driving a one-minute route of the fastest route.}
    \label{fig:divergence_travel_times_happy_route}
\end{minipage}%
\hspace{0.2cm}
\begin{minipage}{.47\textwidth}%.51\textwidth
    \centering
\includegraphics[height=5.28cm,trim=0 .4cm 0 1.5cm]{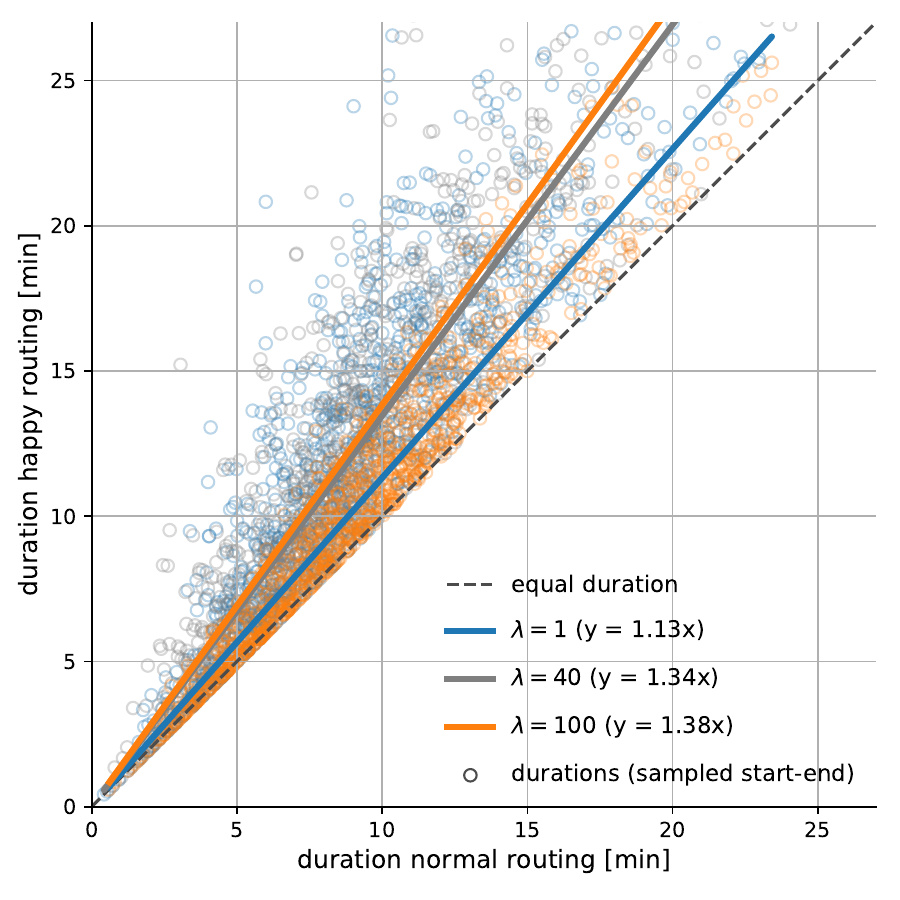}
  \caption{Influence of emotional weighing factor $\lambda$ on happy routes. The additional travel time for happy routing does not scale linearly with the emotional weighing factor $\lambda$. On average, the setting $\lambda=40$ achieves a similar time divergence to $\lambda=100$.}
    \Description[Figure 9 shows how the emotion penalty factor $\lambda$ influences the travel time of the navigation.]{Figure 9 shows how the emotion penalty factor $\lambda$ influences the travel time of the navigation. The scatterplot shows sampled start-end points of durations where the x-axis is the duration of the fastest routing, and the y-axis is the duration when taking the happy route (measured in minutes). Apart from the scatters for different $\lambda$, a regression line indicates the increasing time for happy routing when $\lambda$ increases.}
     \label{fig:lambda}
\end{minipage}
\end{figure*}

We anticipated that taking the happy route would increase the travel time. 
~\autoref{fig:divergence_travel_times_happy_route} shows the relationship of the navigation mode on travel times using $\lambda=20$. 
%In effect, happy routes take longer on most of the rides. 
Using linear regression, we \chreplaced[id=1AC]{found}{find} that a one-minute increase in fastest routing requires in average 1.26 minutes (75.6 sec.) more time to drive using happy routing. 
Only $9\%$ of the start-end coordinates \chreplaced[id=1AC]{resulted}{result} in a situation where the happy route is identical to the fastest route ($overlap = 100\%$). The time difference can be substantial in individual cases. Therefore, we stress a transparent time forecast when recommending happy routes to drivers. We conclude that the factor $\lambda$ should rather be regarded as an internal technical parameter (see the influence of $\lambda$ in~\autoref{fig:lambda}) instead of a user-adjustable parameter. Higher $\lambda$ results in increased average travel time and, therefore, causes longer travel times. Hence, $\lambda$ can be adjusted dynamically to suit the societal driving context.
%We conclude that the factor $\lambda$ should rather be regarded as an internal technical parameter (see influence of $\lambda$ in appendix Figure~\ref{fig:lambda}), instead of a user-adjustable parameter. 
%Notice that we can adapt the time differences of happy routing by changing the factor $\lambda$ for considering emotion weights in the routing optimization (see appendix Figure~\ref{fig:lambda}). 

% \begin{figure*}
% \centering
% \begin{minipage}[b]{.45\textwidth}
% \centering
%  \includegraphics[width=1\linewidth]{figures/results/byalstok_lambda=20_scatter_duration_divergence.pdf}
% \caption{Caption}\label{label-a}
% \end{minipage}\qquad
% \begin{minipage}[b]{.45\textwidth}
% \centering
% \includegraphics[width=1\linewidth]{figures/results/travel_time_divergence_lambda_study.pdf}
% \caption{Caption}\label{label-b}
% \end{minipage}
% \end{figure*}

\subsection{Road Characteristics}
\label{subsec:road_characteristics}
\begin{figure*}[t]
\centering
\begin{minipage}{.47\textwidth}%.45\textwidth
  \centering
    \includegraphics[width=1\linewidth]{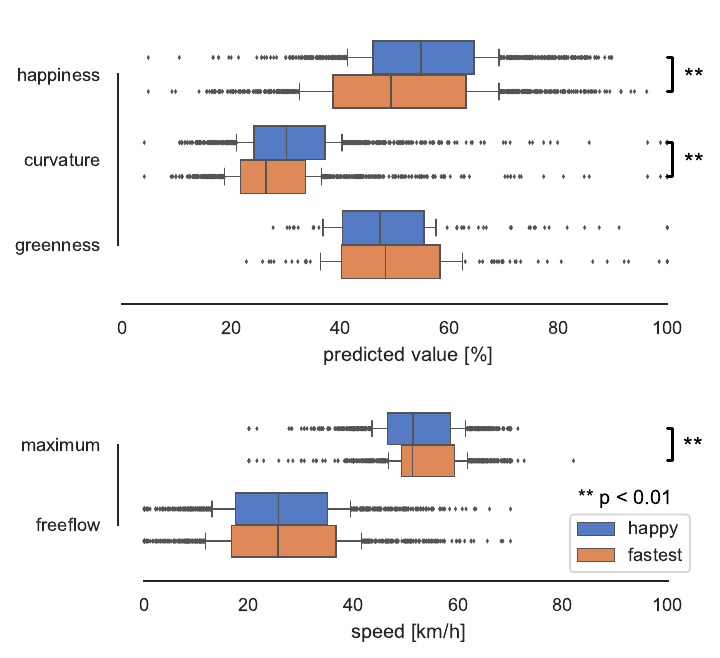}
  \captionof{figure}{Characteristics of happy route vs. fastest route. Distribution of happiness, curviness, greenness, max\_speed, and freeflow\_speed for the two routing modes.}
  \label{fig:numerical_routing_context}
  \Description[Figure 10 shows five bar charts next to each other depicting the distribution of the features for the prediction.]{Figure 10 shows five bar charts next to each other depicting the distribution of the features' happiness', 'curvature', 'greenness', 'max_speed', and 'freeflow_speed'.}
\end{minipage}%
\hspace{0.2cm}
\begin{minipage}{.47\textwidth}%.51\textwidth
  \centering
  \includegraphics[width=1\linewidth]{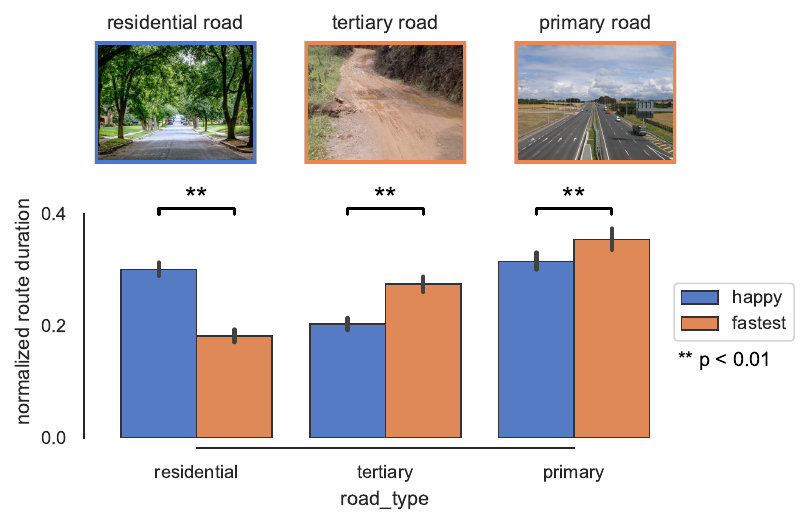}
  \captionof{figure}{Analysis of road types of happy routing vs. fastest routing. We \chreplaced[id=1AC]{assessed}{assess} the road type of every road segment (x-axis) and \chreplaced[id=1AC]{computed}{compute} the drive-time normalized route duration (y-axis). All presented road types have been tested to be significantly different ($p<.01$). Residential roads \chreplaced[id=1AC]{were}{are} found in living areas, primary road types are major highways linking large towns, and  tertiary roads connect minor streets to more major roads.}
  \label{fig:numerical_roadtype_analysis}
  \Description[Figure 11 shows how the different road types influence the road duration.]{Figure 11 shows an analysis of the road types of happy and fastest routing. All road types showed a significant difference for happy and fastest routes for different road types.}
\end{minipage}
\end{figure*}

We \chreplaced[id=1AC]{analyzed}{analyze} the recommended happy and fastest route for their road types \chadded[id=1AC]{with results shown} in~\autoref{fig:numerical_routing_context} and~\autoref{fig:numerical_roadtype_analysis}. 
As the drive-time \chreplaced[id=1AC]{was}{is} normalized per individual route, the values of the bars do not add up to $100\%$. 
We tested whether the distribution of the different road characteristics is significantly different ($p<.01$) using a non-parametric Mann-Whitney U test. 
Compared to the fastest route, we \chreplaced[id=1AC]{found}{find} that happy routes consist of more road segments with a higher predicted happiness score, higher curviness, higher freeflow speed, and maximum speed. 
%We expect that the happy routes have a higher assigned happiness score due to our optimization formula~\ref{eq:routing_opt}. 
%The curviness is computed using a weighted measure of the radial length of curves (for computational details, see Appendix~\ref{appendix:curviness}).

Curvy roads tend to increase driving enjoyment but also inhibit driving accident risks~\cite{haynes2007district}. 
Unhindered traffic scenarios can be captured by our proxy variable free-flow speed, which is higher for happy routes and increases driver well-being~\cite {roidl2014emotional}. 
We \chreplaced[id=1AC]{detected}{detect} no significant effect of the satellite-image-derived greenness (known part of the HSV spectrum) in happy routes compared to the fastest route ($p=.29$). 
Finally, we \chreplaced[id=1AC]{found}{find} that on-average happy routing includes significantly more residential roads. 
\chadded[id=1AC]{We believe that this is due to the fact that} residential roads often have reduced traffic and may reduce drivers' stress, leading to a more happy emotional state. 
In contrast, the recommended fastest routes \chreplaced[id=1AC]{contained}{contain} significantly more living street and primary road segments, which often require more driver attention. As stated before, these findings are based on a large sample size and do not represent an individual recommended route.

\subsection{Computational Characteristics}

Navigation systems deployed in the wild require high scalability. 
To assess the computational complexity of our system, we computed the execution time of the routing endpoint (GraphHopper). 
On the $12 \times 12$ km map, our system needs to perform emotion prediction on $21,673$ unique edges, \chreplaced[id=1AC]{caching}{and caches} the corresponding data in the optimization graph. 
The cache is needed because the input data is collected from various APIs, which makes on-demand prediction attainable when optimizing the route. 
In a subsequent step, the execution time for recommending happy routes is $0.08 \pm 0.075$ seconds and takes longer to compute than the fastest routing $0.01 \pm 0.004$. 
With recommendation times smaller than $1$ second, our system is highly time-efficient and user-friendly. 

\section{Discussion}
\label{sec:discussion}

With \system, drivers perceived a higher valence when using the happy route than the fastest route, showing that choosing emotionally positive routes contributes to a driver's well-being. In the following, we discuss the implications of our results.
%Drivers perceived a higher valence when using the happy route compared to the fastest route, showing that choosing emotionally positive routes contribute to an improved well-being. In the following, we discuss the implications our results

\subsection{Tradeoff Between Valence and Route Duration}
\chadded[id=1AC]{In contrast to previous work~\cite{wang2018SAR}, we empirically evaluated the impact of \system{} in real-world driving scenarios.}\chadded[id=2AC]{} \chadded[id=R3]{} Our results suggest a tradeoff between the duration of the fastest route and the perceived valence of driving the happy route. Although the happy route takes longer, our participants subjectively preferred \system{} for their navigation to improve their emotional well-being. \chadded[id=1AC]{This confirms previous findings regarding the implementation~\cite{wang2018SAR, liu2021empathetic} and user-centered evaluation in laboratory settings~\cite{bethge_vemotion_2021, doi:10.1080/13658816.2014.931585}. In this context, our results are in line with previous research that participants prefer emotional navigation~\cite{doi:10.1080/13658816.2014.931585}} \chadded[id=R3]{}. However, due to the longer travel times, most of our participants indicated that they would prefer the \system{} if they were not pressed for time. In addition, our study results suggest other modalities for controlling driver emotions by combining the in-vehicle environment with the suggested happy route. For example, participants \chreplaced[id=1AC]{suggested}{proposed} to explore music in combination with happy routes to enhance feelings of happiness. Using individual preferences for the in-vehicle environment as an additional variable can lead to emotion prediction models that ultimately reduce driving time. However, combining in-vehicle adaptions with happy routes proposed by \system{} requires further research. 
% R2 specifically asks about tradeoffs between route length/duration and “happiness” and how to optimize the two given that there could be a relationship between the two, but not a well-defined relationship. R1 asks if the study asked users about such tradeoffs explicitly. The paper should also discuss if such tradeoffs could be personalized for each user (R2).
\textcolor{black}{Overall, numerous transportation studies have researched the willingness to pay, i.e., how much time and money people are willing to spend for an alternative route choice. \chreplaced[id=1AC]{The admissible detour duration is highly dependent on the situation and the individual~\cite{calfee1998value}.}{The admissible detour duration is highly situation- and person-dependent~\cite{calfee1998value}.}  Self-centered situations, such as avoiding danger, reach much higher detour acceptance and decline more slowly with longer detours. The user acceptance drops to the $25\%$ plateau at 8-min detours for jam-related situations~\cite{kroller2021driver}.}

\subsection{Benefits of \system{}}
The emotional state of drivers plays a critical role in road safety, as it has been shown that negative emotions, such as anger, can significantly increase accidents~\cite{underwood1999anger}. More specifically, it has been shown by identifying common emotional triggers based on their originating source via driving self-report that the most frequently elicited negative emotions come from the navigation interface of the car. Therefore, car systems proposing appropriate interventions, such as improving the routing choice, are helpful in improving road safety and enhancing the driving experience. Future work will examine the safety reduction potentials of different route choices. By promoting positive emotions through navigation choices, \system{} introduces a novel paradigm in safety-aware routing, reducing emotional distress and fostering a more composed driving experience.

\subsection{Using \system{} for Other Transport Modalities \chadded[id=R2]{and Types}}
%Alles geil machen für andere (cyclists, pedestrians -> sensor is different)
% \todo[inline]{Thomas: schau bitte hier mal über den Text drüber}
\system{} generates routing decisions that can be used in various other transportation modalities once the foundation for a context-aware machine learning classifier is established. With a few modifications, \system{} can apply emotion-based navigation \chreplaced[id=1AC]{, for example for cyclists,}{for cyclists} by predicting emotionally pleasant cycling routes. We propose incorporating advanced contextual sensors when optimizing happy routes for other road users (e.g., pedestrians or cyclists) by extending the feature set to include elevation information and information about road intersections. For the application of \system{} in pedestrian routing, we recommend extending our feature set to include traffic-banning features, as these have been shown to influence valence~\cite{ortag2011location} positively. \chadded[id=1AC]{Such a set of features can extend existing work investigating the relevance of contextual driving features regarding classification accuracy.} \chadded[id=R3]{}

\chadded[id=R2]{Implementing \system{} for autonomous vehicles presents several unique challenges and opportunities. The emotion prediction model may exhibit very different characteristics when the vehicle operates autonomously. Considering the emerging problem of motion sickness in automated vehicles\,\cite{Abhraneil2021MotionSickness}, a traffic jam may be more acceptable in favor of less crowded roads in the countryside. As users may engage in different primary activities, such as working or watching movies, future research must explore the emotional impact on these activities associated with driving.}

% \begin{figure}[h]
%   \centering
%     \includegraphics[width=1\linewidth]{figures/results/happy_vs_fast_analysis_colorblind.pdf}
%   \caption{Characteristics of happy route vs. fastest route. Distribution of happiness, curviness, greeness, max\_speed and freeflow\_speed for the two routing modes.}
%   \label{fig:numerical_routing_context}
%   \Description[Figure 10 shows five bar charts next to each other depicting the distribution of the features for the prediction.]{Figure 10 shows five bar charts next to each other depicting the distribution of the features' happiness', 'curvature', 'greenness', 'max_speed', and 'freeflow_speed'.}
% \end{figure}
% \begin{figure}[h]
% \centering
%   \includegraphics[width=1\linewidth]{figures/results/roadtype_analysis2_colorblind.pdf}
%   \caption{Analysis of road types of happy routing vs. fastest routing. We assess the road type of every road segment (x-axis) and compute the drive-time normalized route duration (y-axis). All the presented road types have been tested to be significantly different ($p<.01$). Residential roads are are found in living areas, primary road types are major highways linking large towns and  tertiary roads connect minor streets to more major roads.}
%   \label{fig:numerical_roadtype_analysis}
%   \Description[Figure 11 shows how the different road types influence the road duration.]{Figure 11 shows an analysis of the roadtypes of happy and fastest routing. All road types showed a significant difference for happy and fastest routes for different road types.}
% \end{figure}

\subsection{Ethics \& Societal Impact}
%\todo[inline]{T: Rename to Ethics \& Societal Impact and include the works from the rebuttal}

We emphasize an ethical and transparent use of \system{} for application purposes and stress that emotions are intimate, personal, and vulnerable~\cite{andalibi2020human}. 
The Emotional Artificial Intelligence ethics guidelines by McStay et al.~\cite{mcstay:ethicsguidelines} provided us with a meaningful reference to cover personal, relationship, and societal aspects. 

Our approach is privacy-aware because it uses a machine learning model based on an aggregate, anonymized dataset provided in advance by a set of volunteers rather than subconsciously assessing the emotions of individual \system{} users. 
On the other hand, we also see clear limitations of our dataset in the area of cultural and regional diversity and the explainability of resulting algorithm choices. 
Future empathic car interfaces must communicate how and what data is assessed to clarify how this subsequently affects the users' privacy. 

Undeniably, the regulation of emotions by technological systems is highly controversial, as psychological effects are largely unknown. Avoidance of negative situations, for example, is an essential strategy of human emotion (self-)regulation~\cite{mcrae2020:emotionegulation}, but also an implicit result of our system's promotion of positive emotions. Studies with individuals have shown that situation avoidance results in decreased learning and adaptation abilities, as well as social and anxiety disorders~\cite{aldao:emotionregulation}.
Therefore, we emphasize that such short- and long-term effects must be investigated in future work. 

%Our study of route characteristics shows that heavily traveled routes are often avoided in favor of quieter routes. 
%To us, this is a clear indication for future work, as these externalities at large-scale can potentially affect residential areas, parks, or nature, as Johnson et al. note~\cite{johnson_externalitites}. 
%We showed that routes proposed by \system{} result in increased travel times which are ultimately bound to higher energy consumption. 

Our study of route characteristics shows that heavily traveled routes are often avoided in favor of quieter routes. To us, this is a clear indication for future work, as these externalities at large scale can potentially affect residential areas, parks, or nature, as Johnson et al. note~\cite{johnson_externalitites}. \textcolor{black}{This raises concerns that optimizing on happiness could further contribute to the frustrations about increased traffic in previously low-trafficked neighborhoods. }

We showed that routes proposed by \system{} result in increased travel times, ultimately bound to higher energy consumption. 
\textcolor{black}{From an environmental standpoint, this higher energy consumption might be harmful. Overall, no ethical guideline would prioritize one's happiness over the negative externalities that may result from their navigation choice~\cite{beauchamp2003methods}. Moreover, prioritizing personal happiness over social and environmental responsibility may also perpetuate a culture of individualism that values personal satisfaction over the greater good. Therefore, it is imperative to consider the broader social and environmental implications of routing decisions, while striving to balance personal happiness and the well-being of others and the planet. }
Certain route choices might affect the safety of traffic participants, for example, due to a model's preference for specific road types. These and many other route characteristics must be communicated transparently to the users to promote their autonomy and enable highly informed choices~\cite{mcstay:ethicsguidelines}. Alternative strategies could comprise correction terms applied to our optimization, for example, when the routing choice is not desirable on a societal basis (e.g., routing through densely populated areas)~\cite{johnson_externalitites}.

\subsection{Limitations \& Future Work}

%\todo[inline]{T: Driver / co-driver dilemma; Travel sickness}

Our work takes the first steps towards a novel type of empathic car interface based on emotional predictions and optimizations through routing. To achieve this goal, we accepted several limitations in the domains of psychology, algorithms, and user experience. First and foremost, the psychological model of fostering well-being through aggregation of positive emotions is deliberately oversimplified, as discussed in Section \ref{sec:concept}. Future models could operate on a diverse emotional flow~\cite{knobloch:moodadjustment}, which requires significant changes to the optimization method and its proven graph algorithms. Yet, the system’s ability to generalize across unseen environments relies on its subject-independent emotion prediction approach. While individual adjustments could further refine classification accuracy, our current implementation provides a scalable solution for real-world deployment.

Utilizing emotion-related signals during driving would enable the dynamic updating of the predicted emotional weights and real-time adaptation of the happy route. This feature can be easily integrated into the current system architecture, but it should be approached with caution as it has the potential to be perceived as privacy-intrusive. However, the benefit of our non-interactive emotion navigation system is that it allows for an empathic interface without compromising privacy during operation, and the option to switch to a different routing modality can be easily selected at any point during the journey.

%Furthermore, having access to emotion-related features during driving the happy route would enable the automatic updating of predicted emotional weights and the real-time adaptation of the happy route. This real-time capability of \system{} is easily extendable in our current framework but should be treated with caution as this always-on emotion-detection is privacy-intrusive. The advantage of this non-interactive emotion navigation system is that it this empathic interface is not privacy-intrusive during driving and its emotion-routing capability can be easily stopped throughout the ride by choosing another routing modality.
%One limitation of this system is that it does not adapt to the driver's emotions in real-time, however, the advantage of this non-interactive emotion navigation system is that it this empathic interface is not privacy-intrusive during driving and its emotion-routing capability can be easily stopped throughout the ride by choosing another routing modality.

\system{} requires the ability to simulate the driver's emotions for any road segment at any time while considering contextual information like traffic, road types, and speed limits. A key design decision for simulation lies in the choice between subjective and objective metrics for characterizing user emotions. \system~relies on a dataset containing self-expressed and thus subjectively perceived emotions for prediction. Consequently, we base the simulated emotions on discrete representations of emotions, as identified by Ekman~\cite{ekman_are_1992}. \chreplaced[id=R2]{The use of subjectively expressed emotion labels could also be accompanied by the integration of physiologically derived labels into the emotion prediction model\,\cite{jansen2013detection}. By incorporating data such as heart rate variability, skin conductance, muscle stiffness~\cite{balters:stresscar}, and facial expressions~\cite{liu2021empathetic}, a hybrid model (objective \& subjective) could further enhance the accuracy of the emotion predictions.}{However, a detailed comparison with objective metrics should be subject to future work, such as using facial expressions~\cite{liu2021empathetic} or muscle stiffness~\cite{balters:stresscar}.} 
\chadded[id=R2]{Related works by Wang et al.\,\cite{wang2018SAR} and Zepf et al.\,\cite{zepf2020empathicgps} highlight the potential of bridging live emotions with future recommendations or an adapted system behavior.} \chadded[id=2AC]{}
At the same time, the car offers only a limited set of remotely accessible contextual features for predicting driver emotions, making the modeling complex. 

Our navigation framework is based on an emotion prediction layer, which can be adapted easily to additional modalities. Weighting in objective parameters such as the greenness score~\cite{baouche2014electric} could promote user-specific preferences without needing a personalized emotion model. On the other hand, user-dependent models can further increase the accuracy, as shown in related work~\cite{bethge_vemotion_2021,balters:stresscar}. 

\chadded[id=1AC]{We acknowledge the sample size limitation (N=13), which impacts statistical power. A follow-up longitudinal study with a larger and more diverse participant pool would strengthen our findings and enable examining long-term behavioral effects of affective navigation. Yet, it is noteworthy that our research represents the first attempt to conduct such a study in a real-world driving environment, including a user evaluation with a functional routing application for unseen roads. We plan to conduct a large-scale study with a more extensive participant pool to address this limitation, incorporating varied routes and study durations. This will be achieved by leveraging crowd-sourcing methods by, for example, distributing \system{} through app stores, thus extending the reach of \system{} to a broader user base and including diversity in our dataset.} \chadded[id=2AC]{} \chadded[id=R3]{}

Finally, we see limitations in explaining the overall recommendation process to the end user, which is ultimately very important for the ethics and transparency of our system. The transparency can lead to placebo effects, where the description of using an allegedly adaptive AI-driven system biases the perceived system utility for drivers~\cite{10.1145/3529225}. In future work, we plan to summarize how route recommendations were computed on an individual user's basis and research how to communicate key emotional route segments~\cite{braun2020if}. 
Finally, further long-term experiments with a larger variety of roads and routes under vastly different conditions are needed to produce necessary evidence of the proposed model's ability to find happy routes. These long-term studies in the wild may help better to understand the effects and societal impact of affective routing.

\section{Conclusion}
\label{sec:conclusion}
This paper presents \system{}, a new type of empathic interface capable of navigating by positive emotions. 
We \chreplaced[id=1AC]{used}{use} personal, environmental, and road-specific information to define a custom emotion routing graph that optimizes routes for happy emotions. This paper validates this novel routing concept through several validations showing external and internal validity. The machine learning classifier used to predict emotion weights has been shown to be able to predict emotions on unseen road elements and driver emotions. Furthermore, a real-world driving study and simulation study \chreplaced[id=1AC]{demonstrated}{demonstrates} its generalizability to extend to unknown routes. Our user study \chreplaced[id=1AC]{showed}{shows} that \system{} elicits positive emotions through navigation. As a consequence, \system{} requires more driving time which was accepted by our participants as long as the circumstances allowed it (e.g., no time pressure). Our work is not only relevant to driving but can also be applied to other areas of mobility and autonomous driving. We are confident that the presented process of simulating emotions and evaluating different paths through many potential user journeys can be generalized to an even wider variety of use cases. 
To encourage the reproducibility of this paper and engage research in this area, we \chreplaced[id=1AC]{published}{publish} the source code of our system and the data set for further analysis by the research community\footnote{\url{https://anonymous.4open.science/r/happyrouting-7FD2/README.md}}.

%Here, by proposing a novel navigation software for emotional routes, we are confident that \system{} advances the field of emotion-aware car interfaces. %To encourage research in this area, we publish the source code and the data set for further analysis by the research
%We make all source code (data, context-emotion classifier, routing engine, mobile navigation app) available online to foster research activity\footnote{\url{missing link}}.

% \begin{acks}
% This work has been partly funded by the German Federal Ministry of Education and Research (BMBF) under Grant No. 01IS18036A (MCML).
% \end{acks}

\bibliographystyle{ACM-Reference-Format}
\bibliography{bibliography}

\appendix
\section{Appendix}
\label{sec:appendix}

%\subsection*{Influence of $\lambda$ on travel time}
%Using the same experimental setting as explained in Section~\ref{sec:simulation}, we analyze the influence of $\lambda$ on travel duration. Figure~\ref{fig:lambda} shows that increasing $\lambda$ results in higher travel time.
% \begin{figure}[H]
%     \centering
%     \includegraphics[width=0.75\linewidth]{figures/results/travel_time_divergence_lambda_study.pdf}
%     \caption{Influence of emotional weighing factor $\lambda$ on happy routes. The additional travel time for happy routing does not scale linearly with the emotional weighing factor $\lambda$. On average, the setting $\lambda=40$ achieves a similar time divergences as $\lambda=100$.}
%     \Description[Figure 11 shows how the emotion penalty factor $\lambda$ influences the travel time of the navogation.]{Figure 11 shows how the emotion penalty factor $\lambda$ influences the travel time of the navigation. The scatterplot shows sampled start-end points of durations where the x-axis is the duration of the fastest routing, and the y-axis is the duration when taking the happy route (measured in minutes). Apart from the scatters for different $\lambda$, a regression line indicates the increasing time for happy routing when $\lambda$ increases.}
%     \label{fig:lambda}
% \end{figure}

\subsection*{Curviness Computation}
\label{appendix:curviness}
We \chreplaced[id=1AC]{computed}{compute} the curviness using a weighted measure of the length of curves, which depends on the radius of a circumscribed circle that passes through all three consecutive geocoordinates in a route. Given $a,b,c$ as the length of the three sides of a triangle, the radius of the circumcircle is given by the formula:
\begin{equation}
    r = \frac{abc}{\sqrt{(a+b+c)(b+c-a)(c+a-b)(a+b-c)}}
    \label{eq:curvature_compute}
\end{equation}

%}

\subsection*{Classifier Feature Importances}
\begin{figure*}[htb]
    \centering
    \includegraphics[width = 0.8\textwidth]{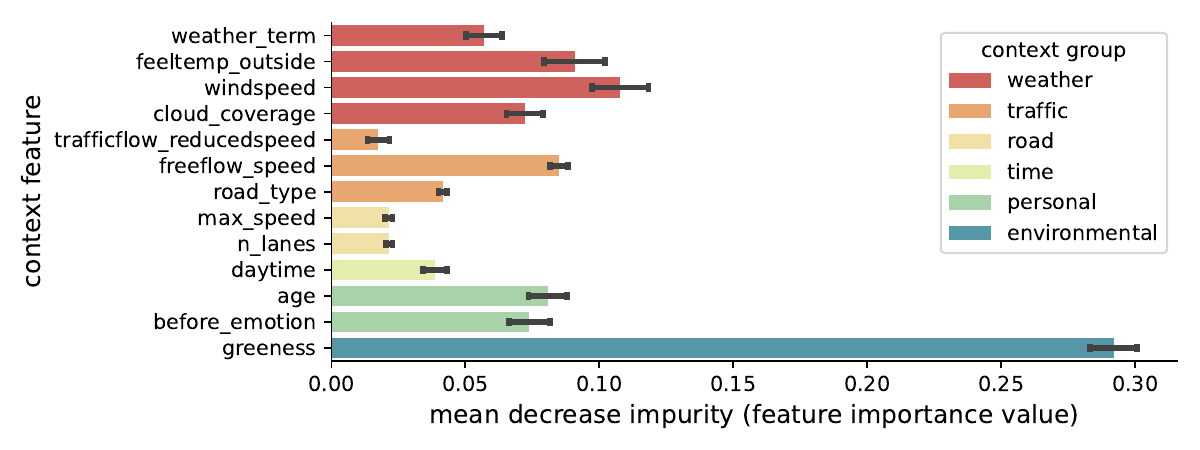}
    \caption{Feature importances measured by the mean decrease of Gini-impurity for the Leave-One-Participant-Out cross-validation.}
    \Description[Figure 12 shows the feature importance of the classifier.]{Figure 12 shows the feature importance of the classifier. On the x-axis the mean decrease of impurity is shown vs. the context features used as input to the classifier (y-axis). Greeness is awarded the highest feature importance. }
    \label{fig:feature_importances}
\end{figure*}

We \chreplaced[id=1AC]{analyzed}{analyze} how decisive each contextual input feature is for our human emotional state classification model. We \chreplaced[id=1AC]{extracted}{extract} the feature importance (Gini impurity) of the input variables in a leave-one-participant-out situation in Figure~\ref{fig:feature_importances}. The variable 'greenness' shows the highest importance for the classifier in predicting the likely emotional state on the road. \textcolor{black}{This likely comes from the fact that roads with high green value scores likely go through rural areas with less traffic flow influencing emotions positively. Thus, 'greenness' is a good proxy input for positive emotional states~\footnote{\textcolor{black}{The training data of the classifier is unbiased, containing green areas and urban road data.}}.  }

The feature importances are aggregate metrics and do not convey participant-dependent importance measures for a specific routing choice (local feature importance measures such as SHAP values are needed). Here, we only \chreplaced[id=1AC]{analyzed}{analyze} the feature importance of the emotion classification model, a route-specific analysis of the road properties can be found in Section~\ref{subsec:road_characteristics}.

\subsection*{In-the-Wild Driving Study}
% \usepackage{booktabs}
% \usepackage{graphicx}
\textcolor{black}{
The in-the-wild driving evaluation routes \chreplaced[id=1AC]{were}{are} $7.5 km$ (fastest) and $8 km$ long and \chreplaced[id=1AC]{went}{go} through urban and rural neighborhoods.
In Table~\ref{tab:questionnaire} we present the study questions used in the in-the-wild driving evaluation of our system. 
}

% \onecolumn

\begin{table*}[htb]
\centering
%\begin{tabularx}{\textwidth}{|r|X|}
\caption{Questionnaire of the in-the-wild driving experiment.}
\label{tab:questionnaire}
\Description[Table with questions of the experiment.]{Table 4 presents the questions used in the in-the-wild experiment.}
\resizebox{\textwidth}{!}{%
\begin{tabular}{@{}ll@{}}
\toprule
Question                                                                                                                                              & Example answer                                                                                                                                 \\ \midrule
What car do you drive?                                                                                                                                & VW Golf                                                                                                                                        \\
Your age                                                                                                                                              & 39                                                                                                                                             \\
Your sex                                                                                                                                              & female                                                                                                                                         \\
How frequent do you drive? (km/year)                                                                                                                  & occassional: <10.000km/year                                                                                                                    \\
When did you drive today?                                                                                                                             & in the morning                                                                                                                                 \\
How do you feel before driving? (valence)                                                                                                             & 3 (of 5)                                                                                                                                       \\
How do you feel before driving? (arousal)                                                                                                             & 5 (of 5)                                                                                                                                       \\
How do you feel before driving? (dominance)                                                                                                           & 4 (of 5)                                                                                                                                       \\
Other notes / suggestions?                                                                                                                            & None                                                                                                                                           \\
What navigation mode did you drive?                                                                                                                   & navigation mode 2                                                                                                                              \\
How do you feel after driving? (valence)                                                                                                              & 2 (of 5)                                                                                                                                       \\
How do you feel after driving? (arousal)                                                                                                              & 4 (of 5)                                                                                                                                       \\
How do you feel after driving? (dominance)                                                                                                            & 4 (of 5)                                                                                                                                       \\
Were there any specific incidence while driving?                                                                                                      & Slow trucks in front of me                                                                                                                     \\
Do you know the route 1? (Have you ever driven this route?)                                                                                           & Partially                                                                                                                                      \\
How would you describe route 1?                                                                                                                       & Slow, green and full of Blitzer                                                                                                                \\
Select all adjectives that in your opinion describe route 1 (select as much adjectives as you want)                                                   & Green, Smooth, Relaxing                                                                                                                        \\
How do you feel before driving? (valence)                                                                                                             & 2 (of 5)                                                                                                                                       \\
How do you feel before driving? (arousal)                                                                                                             & 4 (of 5)                                                                                                                                       \\
How do you feel before driving? (dominance)                                                                                                           & 4 (of 5)                                                                                                                                       \\
What navigation mode did you drive?                                                                                                                   & navigation mode 1                                                                                                                              \\
How do you feel after driving? (valence)                                                                                                              & 3 (of 5)                                                                                                                                       \\
How do you feel after driving? (arousal)                                                                                                              & 3 (of 5)                                                                                                                                       \\
How do you feel after driving? (dominance)                                                                                                            & 3 (of 5)                                                                                                                                       \\
Were there any specific incidence while driving?                                                                                                      & Lkw and trucks in front of me                                                                                                                  \\
Do you know route 2? (Have you ever driven this route?)                                                                                               & Yes                                                                                                                                            \\
How would you describe route 2?                                                                                                                       & A lot of construction work and interruptions                                                                                                   \\
Select all adjectives that in your opinion describe route 2                                                                                           & boring, bumpy, relaxing                                                                                                                        \\
I agree with the following statement: "I feel  route 1 is faster than route 2"                                                                        & equal                                                                                                                                          \\
I agree with the following statement: "I feel  route 2 is faster than route 1"                                                                        & equal                                                                                                                                          \\
Which route would you rather choose?                                                                                                                  & Route 1                                                                                                                                        \\
Why?                                                                                                                                                  & Felt smoother                                                                                                                                  \\
I agree with the following statement: I feel route 1 makes me happier than route 2? (ordered and preprocessed response)                               & yes                                                                                                                                            \\
Do you feel route 2 happier than route 1?                                                                                                             & no                                                                                                                                             \\
Do you feel route 1 happier than route 2?                                                                                                             & yes                                                                                                                                            \\
%If the happy route would be longer, h
How much time you would like to sacrifice to drive a happier route (assuming 20 minutes drive for fastest route)? & 3-5 minutes                                                                                                                                    \\
What did you do apart from driving? Applies to both driving modes                                                                                     & hearing music/radio, talking to passengers                                                                                                     \\
Does something about the Happy Navigation idea bother you?                                                                                            & You drive more                                                                                                                                 \\
When (under which circumstances) would you use Happy Navigation?                                                                                      & When I am somewhere new (e.g., holiday), start the day smoother, when I have more time,  when I want to listen to a podcast \\
What determines your ideal happy driving route (road elements, scenery)?                                                                              & Green, drive by forest, less cars, rapsfelder                                                                                                  \\
When would you use Happy Navigation?                                                                                                                  & In the morning                                                                                                                                 \\
What features in the car would you find interesting using the Happy Navigation?                                                                       & I would want to see if I drive the happy route (transparency)                                                                                  \\
How likely would you use this system?                                                                                                                 & 3 (of 5)                                                                                                                                       \\
Do you think there are any societal and ethical implications of this navigation functionality? And if yes, which one?                                 & Fuel or energy consumption increases, invisibility of unhappy places and roads and their existences, self enforcing effect                     \\
Other notes/suggestions?                                                                                                                              & None                                                                                                                                          \\ 
\bottomrule
\end{tabular}%
}
%\end{tabularx}
\end{table*}

\subsection*{Emotion Classifier Dataset}
\label{subsec:appendix_classifier_dataset_description}

%\todo[inline]{It should also contain the demographics of the dataset used for the emotion prediction model (age groups, culture, countries) (R2). Should contain more information on how the dataset was obtained (R3)}

% comment= It should also contain the demographics of the dataset used for the emotion prediction model (age groups, culture, countries) (R2). Should contain more information on how the dataset was obtained (R3)}
% chadded does not work for paragraphs!
%\chadded[id=R3]{
%The dataset for training the machine learning model for predicting emotional weights of the routing map layer is described below.

%describe the study design
\paragraph{Dataset Generation Procedure}
\chadded[id=1AC]{
An iOS app was developed to track GPS and video during car rides. The smartphone app gathered information about the road type, greenness, traffic flow, and other variables to describe the driving context. Upon hearing a beep, participants were asked to self-report their emotions every 60 seconds. The participant's self-reported emotions were the ground truth in this real-world experiment. The timing of these prompts was fine-tuned in a small pre-study to ensure safety and minimize distraction. Most participants found these prompts non-disruptive, responding within an average of 1.8 seconds. Additionally, participants were provided with a list of basic emotions before the experiment.
Participants were drawn from a group of willing colleagues contacted via a mailing list, prepared by downloading our iOS app and securing a windshield smartphone holder. Before their next drive, they engaged in a remote conversation with the study instructor, sharing demographics, driving habits, and pre-ride emotions. Following an introduction to the app, they commenced recording, drove to their destination, saved recordings post-ride, and subsequently connected with the instructor. During this call, they discussed noteworthy driving incidents and their emotional experiences before, during, and after driving. Ethical approval was granted by the institutional review board of the university department.}\chadded[id=R3]{}

\paragraph{Emotion Labeling}
\chreplaced[id=1AC]{For the emotion classifier, we leverage existing approaches that link real-world context and emotions~\cite{bethge_vemotion_2021,bethge2022designspace}. The authors built an iOS app to record GPS and video during car rides and compute variables continuously. Participants were asked to use this app and attach their phones to the windscreen during their next car ride. The authors recorded the daytime and participants' emotions at the ride's beginning. A beep tone was triggered every $6$ seconds for participants to verbally provide their currently perceived emotions. This emotion probing corresponds to the \textit{in-situ} categorical emotion response (CER) rating for collecting data on emotional experiences in vehicles~\cite{dittrich2019exploring}. Before starting the experiment, participants were instructed about the set of available emotions (i.e., Ekman's basic emotions~\cite{ekman1984expression}). The verbally expressed emotions were recorded and analyzed afterward using a speech-to-text algorithm. In a pre-study ($N=5$), the time interval of the prompts was optimized to ensure safety, minimize annoyance, and appropriately cover the felt emotions. Participants found these prompts non-disruptive, responding within an average of $1.8$ seconds.}{For our emotion classifier for unknown routes, we leverage existing approaches that performed real-world context-emotion linking, in particular \cite{bethge_vemotion_2021} and \cite{bethge2022designspace}.
The authors built a vehicle-usable iOS app that records the individual GPS and video stream and computes the variables continuously during the ride. The participants were asked to use this app the next time they used their personal car to ride and attach their phones to the windscreen. 
The authors recorded the daytime and asked the participants about their emotions at the ride's beginning. To collect a baseline of the participant's own interpretation of emotional states during the ride, a beep tone is triggered every 60 seconds for the participants to verbally provide their currently perceived emotions. The emotion probing is designed in correspondence to the \textit{in-situ} categorical emotion response (CER) rating for collecting data on emotional experiences in vehicles~\cite{dittrich2019exploring}. Participants were instructed about the available emotions before starting the experiment (i.e., the basic emotions after Ekman~\cite{ekman1984expression}). The verbal emotion was recorded while driving and analyzed after the driving scenarios with a speech-to-text algorithm. As this procedure requires the passenger to talk during the ride and can be a distraction from first-order driving tasks, in a pre-study ($N=5$), the time interval is optimized not to be annoying, ensure safety, and simultaneously cover the felt emotions appropriately. For an in-the-wild system that uses our architecture, the ground truth emotion assessment will not be required, and therefore, the system will not interact with the driver. A printout of the basic emotions was given to the participants before the start of the experiment.}

\paragraph{Dataset Description}
\chadded[id=1AC]{
The dataset contains 26 participants (17 male and 9 female) with an average age of 30 years (SD = 5.56) from Germany, Brazil, and Poland. Most driving sessions (69\%) occurred in Germany, followed by Brazil (23\%) and Poland (9\%). Participants reported driving frequently (57\% over 30,000 km/year), moderately (19\% between 10,000 and 20,000 km/year), or less frequently (23\% below 10,000 km/year). Driving sessions included rural and urban roads, lasting an average of 24 minutes (SD = 12, min = 7, max = 52) with 5.6 changes in road type per ride. Participants typically expressed 5.21 distinct emotions during their rides, with an average time of 2 minutes (SD = 2.5 minutes) between each expressed emotion.}\chadded[id=R3]{}

\end{document}